\pdfoutput=1
\documentclass[aip, reprint, superscriptaddress, letterpaper, floatfix]{revtex4-2}
\usepackage{graphicx}
\usepackage{upgreek}
\usepackage{color}
\usepackage{subfigure}
\usepackage{tabularx}
\usepackage{floatrow}
\usepackage{dcolumn}
\usepackage{bm}
\usepackage{amsmath}
\usepackage{scrextend}
\usepackage[normalem]{ulem}
\usepackage{hyperref}

\begin{document}

\title{Terahertz frequency-domain $4\times4$ Mueller matrix ellipsometer instrument designed for high-frequency magnetic resonance measurements}

\author{Viktor Rindert}
\thanks{These authors contributed equally to this work}
\email[Electronic mail: ]{viktor.rindert@ftf.lth.se}
\affiliation{NanoLund and Solid State Physics, Lund University, S-22100 Lund, Sweden}
\affiliation{Terahertz Materials Analysis Center, THeMAC, Lund University, S-22100 Lund, Sweden}
\affiliation{Center for III-Nitride Technology, C3NiT - Janz\'en, Lund University, S-22100 Lund, Sweden}

\author{Alexander Ruder}
\thanks{These authors contributed equally to this work}
\altaffiliation[Current address: ]{Onto Innovation Inc., New Jersey, USA.}
\affiliation{Department of Electrical and Computer Engineering, University of Nebraska-Lincoln, Lincoln, NE 68588, USA}

\author{Steffen Richter}
\altaffiliation[Current address: ]{Schott AG, Mainz, Germany}
\affiliation{NanoLund and Solid State Physics, Lund University, S-22100 Lund, Sweden}
\affiliation{Terahertz Materials Analysis Center, THeMAC, Lund University, S-22100 Lund, Sweden}

\author{Philipp Kühne}
\affiliation{Department of Physics, Chemistry, and Biology (IFM), Link{\"o}ping University, SE 58183, Link{\"o}ping, Sweden}
\affiliation{Terahertz Materials Analysis Center, THeMAC, Link{\"o}ping University, SE 58183, Link{\"o}ping, Sweden}
\affiliation{Center for III-Nitride Technology, C3NiT - Janz\'en, Link{\"o}ping University, SE 58183, Link{\"o}ping, Sweden}

\author{Mark Bauer}
\affiliation{Department of Electrical and Computer Engineering, University of Nebraska-Lincoln, Lincoln, NE 68588, USA}

\author{Vanya~Darakchieva}
\email{vanya.darakchieva@ftf.lth.se}
\affiliation{NanoLund and Solid State Physics, Lund University, S-22100 Lund, Sweden}
\affiliation{Terahertz Materials Analysis Center, THeMAC, Lund University, S-22100 Lund, Sweden}
\affiliation{Center for III-Nitride Technology, C3NiT - Janz\'en, Lund University, S-22100 Lund, Sweden}
\affiliation{Department of Physics, Chemistry, and Biology (IFM), Link{\"o}ping University, SE 58183, Link{\"o}ping, Sweden}

\author{Mathias Schubert}
\email{mschubert4@unl.edu}
\affiliation{NanoLund and Solid State Physics, Lund University, S-22100 Lund, Sweden}
\affiliation{Terahertz Materials Analysis Center, THeMAC, Lund University, S-22100 Lund, Sweden}
\affiliation{Department of Electrical and Computer Engineering, University of Nebraska-Lincoln, Lincoln, NE 68588, USA}

\date{\today}

\begin{abstract}
We report a Mueller matrix ellipsometer design using dual continuously rotating anisotropic meta wave plates which determines the full set of Mueller matrix elements in the terahertz spectral range. The instrument operates in the frequency domain and employs a frequency tunable, solid state synthesizer based, continuous wave terahertz source with sub-MHz bandwidth. The implemented source permits operation within {82-125~GHz and 170-250~GHz, without and with an additional frequency doubler, respectively.} The terahertz transparent meta wave plates consist of {3D-}printed polymer based columnar thin film structures with subwavelength dimensions. The rotating wave plates produce sufficient modulation of the Stokes vector components of transmitted terahertz light to use the wave plates in polarization state generator and polarization state detector devices. Fast terahertz light detection rate of a quasi optical solid state detector permits acquisition with few microsecond {temporal} resolution and electronic sweeping control of the source frequency. We develop a fast {frequency sweeping} scheme while continuously rotating the terahertz wave plates. Subsequent sorting of measured data permits measurements of Mueller matrix elements at hundreds of different wavelengths. The Mueller matrix elements are obtained by forward numerical reduction of the measured data using an algorithm developed by Ruder~\textit{et al.}~[Opt.~Lett.~45,~3541~(2020)]. The instrument is combined with a magnetocryostat capable of reaching magnetic field strengths at the sample position of -8~T$\dots$8T and sample temperatures from 4~K$\dots$400~K. Hence, the instrument is suitable to measure the full Mueller matrix of samples with magnetic resonances as demonstrated recently by Rindert~\textit{et al.} [Phys.~Rev.~B~110,~054413~(2024)]. We discuss design, calibration, and example applications.   
\end{abstract}

\maketitle
\section{Introduction}
Ellipsometry measures the change in the polarization state of light after it interacts with a sample. This change can be measured with {high} accuracy and precision, provided the ellipsometer instrument is properly designed, calibrated, and operated. The polarization shift can arise from various physical origins, including the geometry and composition of surfaces or multilayered structures, as well as the initial polarization state of the incident light, which can vary infinitely. To manage this complexity, an {ellipsometer} condenses the information into a more manageable form, often represented by Jones or Mueller matrix elements. These matrices describe the overall effect of a sample on any incident polarization state. Through appropriate modeling, the matrix elements can be used to derive specific material properties, such as the dielectric function or magnetic susceptibility, including their directional (tensor) and frequency dependent (spectroscopic) characteristics.\cite{Fujiwara} Unlike intensity only methods, ellipsometry thereby provides a more detailed analysis by incorporating polarization modulation, offering deeper insights into material properties.

Spectroscopic Ellipsometry is the preeminent ellipsometric method for characterizing the dielectric properties of bulk materials and thin films. Its sensitivity to ultra thin layers and its ability to distinguish between multiple layers in complex heterostructures have made it indispensable in modern research and advanced manufacturing. Improvements in ellipsometry instrumentation address faster data acquisition and {extending} accessible spectral regions. Likewise, expanding spatial scope by developing imaging and scanning instrumentation with increasing spatial resolution is of contemporary interest. Advances also include access to the entire set of 16 real valued elements of the Mueller matrix, which requires the incorporation of dual sets of polarization modulation devices capable of shifting the {phase difference between the two different polarization components}                     of an arbitrarily polarized light beam. Such instruments have been available for over half a century in the visible to ultraviolet regions of the electromagnetic spectrum,\cite{Hauge:78} and have gradually extended into the far infrared range.\cite{10.1063/1.4789495} Extending Mueller matrix ellipsometry into the terahertz region, however, has been particularly challenging due to the "terahertz gap"—a frequency range where the development of optical components has lagged behind other spectral regions.\cite{PAWAR2013157} 

Nevertheless, recent advancements are helping to close this gap. A wide range of optical components necessary for constructing functional terahertz full Mueller matrix ellipsometers are now available, including wire grid polarizers made from carbon nanotubes,\cite{doi:10.1021/nl900815s} aluminum,\cite{Ferraro:16} and thin films.\cite{6357219} Similarly, advancements in waveplates, crucial for modulating the polarization state of light, have been achieved through liquid crystals,\cite{10.1063/1.3626560} highly birefringent wood,\cite{Reid:06, 6119245} and terahertz meta waveplates.\cite{Hernandez, Rohrbach, perspective} Use of the latter is presented in this study. In terms of terahertz sources, backward wave oscillators (BWOs), and solid state sources such as synthesizers and voltage controlled oscillators (VCO) with frequency multipliers have been employed to achieve high resolution spectroscopic ellipsometry measurements. Terahertz ellipsometry can be categorized into frequency domain terahertz ellipsometry (FD-THzE) and time domain ellipsometry (TD-THzE).\cite{10.1063/5.0094056} The fundamental differences between these approaches necessitate distinct instrumentation for FD-THzE and TD-THzE.

The TD-THzE instruments are more widespread due to the ready availability of short pulse high power laser systems. Most state of the art TD-THzE systems utilize photoconductive antennas for both the detector and the source, which are well suited for pulsed terahertz generation and detection due to their broad bandwidth and fast response times.\cite{8735790, app12083744, Mazaheri2022} TD-THzE instruments permit the acquisition of ellipsometry data in a wide spectral range, typically covering up to a few THz. However, the resolution is on the order of a few GHz{, which is insufficient for the purposes presented here}. Instruments capable of acquiring the full Mueller matrix are not readily available.\cite{Neshat:12, Matsumoto:11, 10.1063/5.0094056, Xu:20,10.1063/1.1426258, Mazaheri:22, Agulto:22} For magnetic resonance measurements performed using terahertz spectroscopic ellipsometry, narrow spectral resolution is crucial, as resonance peaks typically have full width half maxima on the order of MHz.\cite{10.1063/5.0082353} Furthermore, coupling of electron spin with its nuclear or nearby nuclear spins causes hyperfine or superhyperfine structures, respectively, with resonance splittings in the sub MHz range. Hence, spectroscopic detection of paramagnetic resonance in the terahertz range requires instrumentation with well defined frequency and narrow bandwidth operation. A fast scanning, fast detection FD-THzE instrument would then be ideal.

In FD-THzE, the typical setup involves a rotating linear polarizer serving as the analyzer, paired with either a fixed or rotating linear polarizer as the {polarization state} generator. Historically, BWO sources have been employed, often in conjunction with Golay cells as detectors.\cite{10.1063/1.4889920,10.1063/1.3297902,7534823, 2018LuEllipsometer,HOFMANN20112593} However, the inherent limitations of Golay cells, particularly their low sampling frequencies in the few Hz range, have driven the adoption of quasi optical detectors (QOD) in more recent FD-THzE setups.\cite{Belyaeva2021-in} QOD detectors offer significantly higher {readout} rates, making them ideal for fast FD-THzE applications. Despite this advantage, quasi optical detectors exhibit reduced sensitivity to terahertz radiation {with increasing frequency}, which can limit their effectiveness in certain measurements.\cite{QOD1, QOD2}

{Electron Paramagnetic Resonance (EPR) is a ubiquitous electromagnetic technique that is highly effective in probing spin densities in materials containing unpaired electrons. The simplest example of EPR is that under exposure to a magnetic field, an unpaired {s-}electron aligns its spin with or against the field, resulting in two distinct energy levels known as Zeeman splitting. The energy separation between these levels is directly proportional to the magnetic field strength. In a standard EPR experiment, a sample is positioned within a waveguide cavity and exposed to a magnetic field.\cite{doi:10.1142/3624, WeilBolton2007} Simultaneously, the waveguide is excited with electromagnetic radiation. {At $B = 357$~mT, the resonance occurs at approximately 10~GHz for a free electron, where the magnetic field strength is given by $B = \frac{\Delta E}{g\mu_{B}}$, with $\mu_B = \frac{e\hbar}{2m_e}$ being the Bohr magneton, $e$ the unit electric charge, $\hbar$ the reduced Planck constant, and $m_e$ the free electron mass.} The associated vacuum wavelength is $\lambda \approx3$~cm. Due to the geometric restrictions of the resonance cavity it is {not possible} to continuously vary the resonance frequency {over a wide range} and thereby the test frequency dependence of the EPR experiment. Therefore, most EPR instruments vary the magnetic field. Likewise, modulation of the polarization is difficult, if not impossible, due to the resonance condition requirement and the necessary modifications of the near field cavity geometry. To improve sensitivity to absorbance loss across spin transitions, an alternating current in an adjacent coil induces a minor fluctuation in the magnetic field at the sample, causing variations in absorption as the energy levels of the unpaired electrons are modulated.\cite{al1964electron,höfer}}

{As an alternative to magnetic field swept EPR, frequency domain EPR (FD-EPR) spectroscopy was developed in the 1960s by Richards \textit{et al.},\cite{Richards1,Richards2} with notable advancements in the 1980s led by Sievers and Richards.\cite{Sievers} However, widespread development of FD-EPR only began in the early 2000s, driven by advances in terahertz technology.\cite{B305328H} Since then, multiple research groups have created high frequency FD-EPR instruments, as outlined in Ref.~\onlinecite{HFEPR}. Of particular note, Schnegg \textit{et al.}\cite{Schnegg} developed a terahertz FD-EPR Fourier transform instrument, Neugebauer \textit{et al.}\cite{Neugebauer} built an ultra broadband EPR spectrometer, and Del Barco \textit{et al.}\cite{Barco} introduced a terahertz EPR spectrometer utilizing a double Martin–Puplett interferometry approach. The aforementioned instruments offer several advantages over conventional magnetic field swept EPR. Notably, sweeping the frequency is considerably faster than ramping a magnet, {and enables } two dimensional magnetic field versus frequency measurements, commonly known as Zeeman diagrams. This capability is particularly useful for accurately determining spin Hamiltonian parameters, which generally require multi frequency measurements. Additionally, the ability to swiftly scan across wide frequency ranges is highly beneficial for samples with magnetic resonance spectra spanning a broad range. Furthermore, a high frequency approach is essential for samples exhibiting zero field splitting or crystal field splitting on the order of hundreds of GHz. Similar capabilities are also present in high frequency time domain EPR instruments.\cite{tdsepr1,tdsepr2}}

{{Higher frequencies are necessary as} far field measurements of the magnetic resonance{s} {using conventional X- (10~GHz) or Q-band (40~GHz) frequencies} would require extremely large samples and quasi optical setups to reach sufficient far field conditions, i.e., when the separation between source and sample as well as sample dimensions must be large against the vacuum wavelength.\cite{JacksonCED} On the other hand, for example, at 20 fold the magnetic field {used in X-band EPR}, the resonance {of a free electron} occurs at approximately 200~GHz and approximately 7~T, then with associated wavelengths at approximately 1.5~mm. Hence, optical instrumentation can be designed to maintain the principle of far field operation. Flat surfaced samples with $>10$~mm diameter can be used to direct incident polarized terahertz radiation towards its surface, and the change of polarization of either reflected and/or transmitted beams can be analyzed following principles of ellipsometry at terahertz frequencies, as demonstrated previously.\cite{doi:10.1063/1.4889920,2018LuEllipsometer} Due to the intrinsic anisotropic nature of magnetic resonance, the response of a sample due to magnetic resonance causes polarization mode conversation.\cite{Schubert04,Schubert:16,Schubert:03,Schubert:96}}

{We propose high frequency electron paramagnetic generalized spectroscopic ellipsometry (HFEPR-GSE) as an advanced far field technique and an improvement over previous FD-EPR instruments. In HFEPR-GSE, a sample is placed within a superconducting magnet capable of generating a static magnetic field in the 0-8~T range, and a subsequent terahertz spectroscopic ellipsometry measurement is performed. Accordingly, electromagnetic radiation with varying polarization is directed onto the sample, and the resulting reflected or transmitted radiation is detected. Unlike EPR methods that use an auxiliary coil to alter the magnetic field for field derivative intensity measurements, HFEPR-GSE uses polarization modulation to determine the polarization response caused by magnetic resonance. {In} HFEPR-GSE {one} can then freely {scan} the measurement frequency within the operational spectral range of the instrument, which, in principle, is not limited by any fundamental restrictions such as cavity resonance conditions. The spectral measurements can then be repeated as a function of the magnetic field. A true two dimensional measurement of the full polarization dependent response of magnetic resonance as a function of energy and magnetic field can be performed.}

{In this study, we present an advanced FD-THzE instrument and its measurement scheme that significantly enhances the capabilities for HFEPR-GSE measurements when combined with a superconducting magnet. The instrument continuously modulates the polarization state and source frequency. A custom synthesizer generates a controllable frequency in the {microwave} range, upconverted by a {cascade of} frequency multipliers. Depending on further combinations of multipliers, the range from approximately 0.1~THz to approximately 1~THz can be covered, although not all of this range is demonstrated here. Detection is performed by a QOD with a maximum response rate of approximately 1~MHz. Continuous polarization state modulation is achieved through rotating terahertz meta waveplates made from 3D printed {metagratings.}\cite{Hernandez, Rohrbach} This setup accelerates data acquisition, provides all Mueller matrix elements, and improves the resolution and accuracy of HFEPR-GSE measurements, representing a significant advancement in the field. Our paper comprehensively describes the instrument's optical, optomechanical, and electronic hardware components, and presents the calibration methodology. We begin by detailing the optical components, their configuration, and the rationale behind their arrangement. Subsequently, we discuss the data acquisition hardware, focusing on synchronizing frequency modulation, polarization state modulation, and analog to digital conversion. The calibration algorithm, including calibration samples and example calibration results, is outlined. We then present examples of experimental measurements and analyses, including an optical model of an $m$-plane sapphire substrate, a HFEPR-GSE measurement of a nitrogen doped SiC sample, and a Zeeman diagram obtained from a Cr doped $\beta$-Ga$_2$O$_3$ sample. The examples demonstrate the versatility and accuracy of our instrument and methodology.}
\section{Instrument description}

\begin{figure}
    \centering
    \includegraphics[width=1\linewidth]{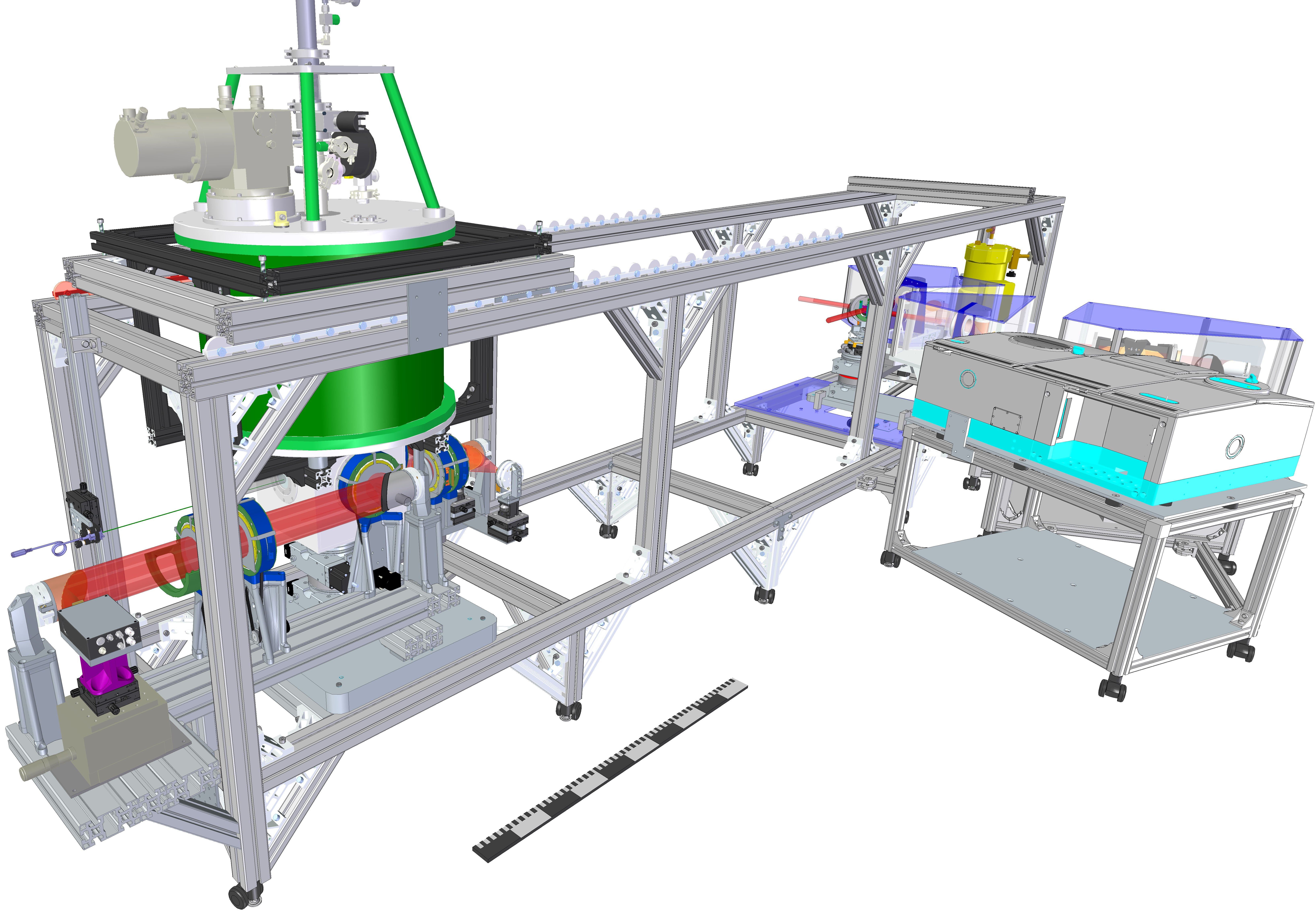}
    \caption{Technical drawing (aerial view to scale, scale bar is 1~m) of the terahertz ellipsometer with splitcoil superconducting magnet and the previously built Optical Hall effect instrument to the right at the THz Materials Analysis Center, see Refs.~\onlinecite{KnightRSI2020,2018LuEllipsometer,doi:10.1063/1.4889920}. The new instrument is shown on the left. The splitcoil superconducting magnet is mounted on a rail allowing it to be moved between the two instruments.}
    \label{fig:OHE+EPR}
\end{figure}

\begin{figure}
    \centering
    \includegraphics[width=1\linewidth]{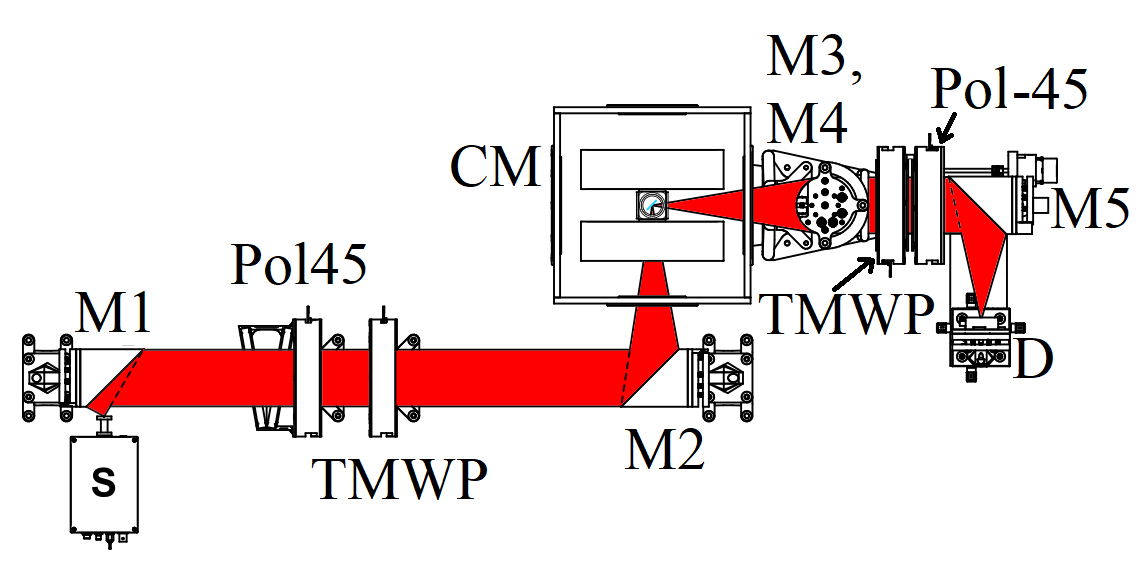}
    \caption{Technical drawing (in true proportions; top view) of the terahertz ellipsometer optical assembly. A splitcoil superconducting magnet (CM) capable of generating magnetic fields from -8~T to 8~T houses the sample under ultra high vacuum conditions and controlled temperature from approximately 4~K to {400~K}. The single frequency, frequency tunable, continuous wave,  synthesizer controlled source (S) consisting of a synthesizer and extension signal generator combination, is operated for the examples shown in this work in the 170-250~GHz range. The terahertz beam originating from the source is collimated by an off axis parabolic mirror (M1), directed through a wire grid polarizer (P45), a terahertz meta wave plate (TMWP), and focused onto the sample inside CM through a pair of outer and inner windows by another off axis parabolic mirror (M2). The beam reflected from the samples through another inner and outer window pair under a 90$^{\circ}$ angle is collected and redirected into a collimated beam by a parabolic mirror (M3), flat mirror (M4) through another terahertz meta wave plate (TMWP), another wire grid polarizer (P-45), redirected and refocused onto a quasi optical detector (D) using an off axis parabolic mirror. P45, P-45, and both TMWPs are mounted onto belt driven rotation stages. The magnetic field orientation is in this {horizontal} plane through the center of the sample and along the incident beam direction. The setup can also be aligned such that the beam entering the sample is collected in transmission at the opposite end of CM by translating the detector stage accordingly. All elements on the input side prior to the beam entering the magnet windows are mounted onto a base plate which is connected with a double ($\theta-2\theta$) goniometer. All elements on the output side after the beam leaving the magnet are mounted onto the first of the goniometer stages. The second goniometer stage is used to mount a sample outside the magnet for calibration purposes with the magnet removed. See also Fig.~\ref{fig:detector_side} for details of the detector stage.}
    \label{fig:instrument}
\end{figure}

\begin{figure}
    \centering
    \includegraphics[width=0.8\linewidth]{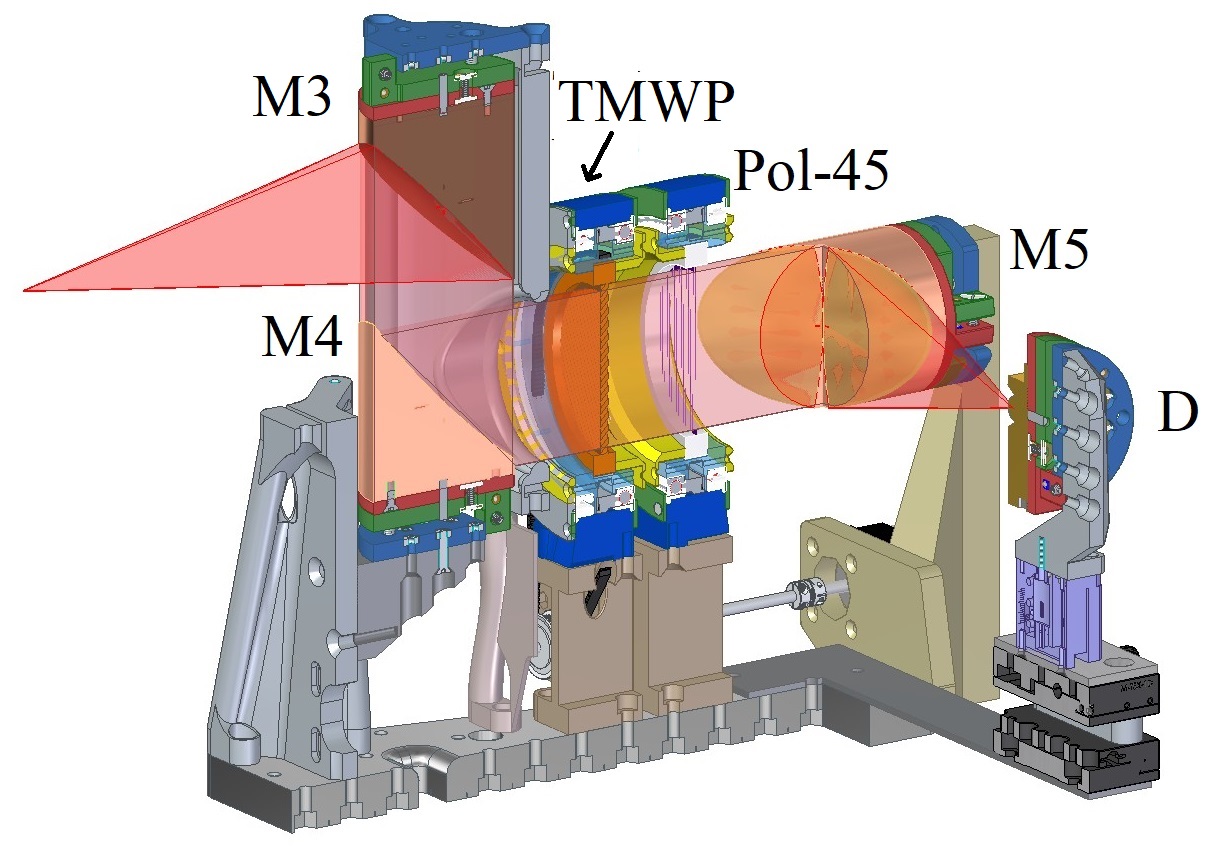}
    \caption{The optical path in the instrument on the detector side. The terahertz beam is marked in red, here originating to the left from the sample (not shown). 90$^{\circ}$ off axis parabolical mirror (M3) collimates and directs the beam onto flat mirror (M4), which redirects the beam through a terahertz meta waveplate (TMWP) and linear polarizer (P-45) onto 90$^{\circ}$ off axis parabolical mirror (M3), which focuses the beam onto the detector (D). {The }TMWP is mounted onto {a} belt driven rotation stage. All elements are mounted onto a stage shown here, which is mounted onto a double rotation stage ($\theta-2\theta$ goniometer; not shown).}
    \label{fig:detector_side}
\end{figure}

\subsection{Terahertz ellipsometer optical assembly}
The terahertz ellipsometer optical assembly is part of a larger optical setup which includes a terahertz Optical Hall effect instrument described previously,\cite{2018LuEllipsometer,doi:10.1063/1.4889920} as shown in Fig.~\ref{fig:OHE+EPR}. The terahertz ellipsometer optical assembly is shown in Fig.~\ref{fig:instrument}, and details of the detector side are shown in Fig.~\ref{fig:detector_side}. The terahertz beam is drawn as red. All mirrors {except} M4 are gold plated and are marked as M1 to M5. The source is labeled {S}, the polarizers as P45 and P-45, the terahertz meta waveplates (TMWP) as TMWP, the cryostat magnet as CM, and the detector as D. The assembly is characterized by its terahertz optical path and all critical components for instrument operation. These include the frequency tunable source, P45 and P-45, two {continuously rotating} TMWPs mounted on separate rotation stages, a detector, and electronic data acquisition timing and control board. {The apertures of the TMWPs and linear polarizers are 8~cm.} Further required for operation as a terahertz ellipsometer suitable for HFEPR-GSE is a superconducting magnet with optical access to the sample placed within a homogeneous section of the magnetic field generated by a splitcoil, i.e., Helmholtz magnet setup. Source, P45, and first TMWP constitute the polarization stage generator (PSG) and are mounted as one fixed unit. Second TMWP, P-45, and detector constitute the polarization state detector (PSD) and are mounted as one separate unit. {The} PSD is mounted onto a goniometer such that the angle of incidence onto the sample can be varied when the magnet is removed. All components are described in detail below.

\subsection{Frequency tunable terahertz Source}
The terahertz source consists of a computer controlled synthesizer that emits electromagnetic radiation in the frequency range of 9.1-13.9~GHz, which is subsequently multiplied by a factor of 9 using a signal generator extension (SGX) module ({synthesizer and SGX made by} Virginia Diodes,~Inc., {VA, USA}), resulting in a usable output range of 82-125~GHz. The {synthesizer} bandwidth is approximately 50~kHz. A frequency doubler {can be} connected to the SGX, extending the output range to 170-250~GHz. The synthesizer is computer controlled, allowing for rapid frequency sweeping across the 170-250~GHz range within 150~ms. The source emits polarized light and is mounted at 45$^\circ$ azimuthal rotation to maximize transmission through the first polarizer. In principle, using a frequency tripler, and combining doubler and tripler, the accessible spectral range can be extended to 246-375~GHz, and 492-750~GHz, respectively. {However, the sensitivity of the QOD utilized in this study decreases with frequency, requiring the use of another detector, such as a bolometer, for these ranges.}\cite{VDI} 

\subsection{Broadband terahertz detector}
To detect the terahertz radiation a quasi optical detector (D) is used. The quasi optical detector {has a maximum read out rate of} approximately 1~MHz, making it well suited for our fast frequency sweeping instrument operation as detailed below. The quasi optical detector consists of a zero biased Schottky diode {connected to a logspiral antenna (ACST GmbH, Hanau, Germany). The log spiral antenna enables a broad frequency range to be detected, independent of the polarization state, and the light is focused onto it by a Silicon lens. The diode is also connected to a{n internal} preamplifier, which is the limiting factor for the readout rate.} The detector has a sensitivity of 20000~V/W at 170~GHz and 14000~V/W at 250~GHz, with a noise equivalent power of 6~pW/Hz$^{1/2}$.

\subsection{Linear polarizers}
Two linear polarizers (P45, P-45) composed of parallel wire grids control the polarization state of the terahertz radiation. The wire grid polarizers are fabricated from 5~$\mu$m diameter tungsten wires (Purewavepolarizers Ltd., Coventry, UK). The wire spacing is 16~$\mu$m, and the {extinction ratio is at least 100:1 in the frequency range 170-250~GHz}. The wire grid polarizers are mounted into manually adjustable, full circle rotation stages. All components of the ball bearing supported rotation stages are manufactured from non ferromagnetic materials. {To reduce standing waves, the polarizers are also rotated 10 degrees around the propagation plane normal. This method has been proven effective in previous studies for standing wave suppression and was therefore implemented here.\cite{2018LuEllipsometer}}    

\subsection{Terahertz meta wave plate}
\begin{figure}[!tbp]
    \centering
    \includegraphics[width=1\linewidth]{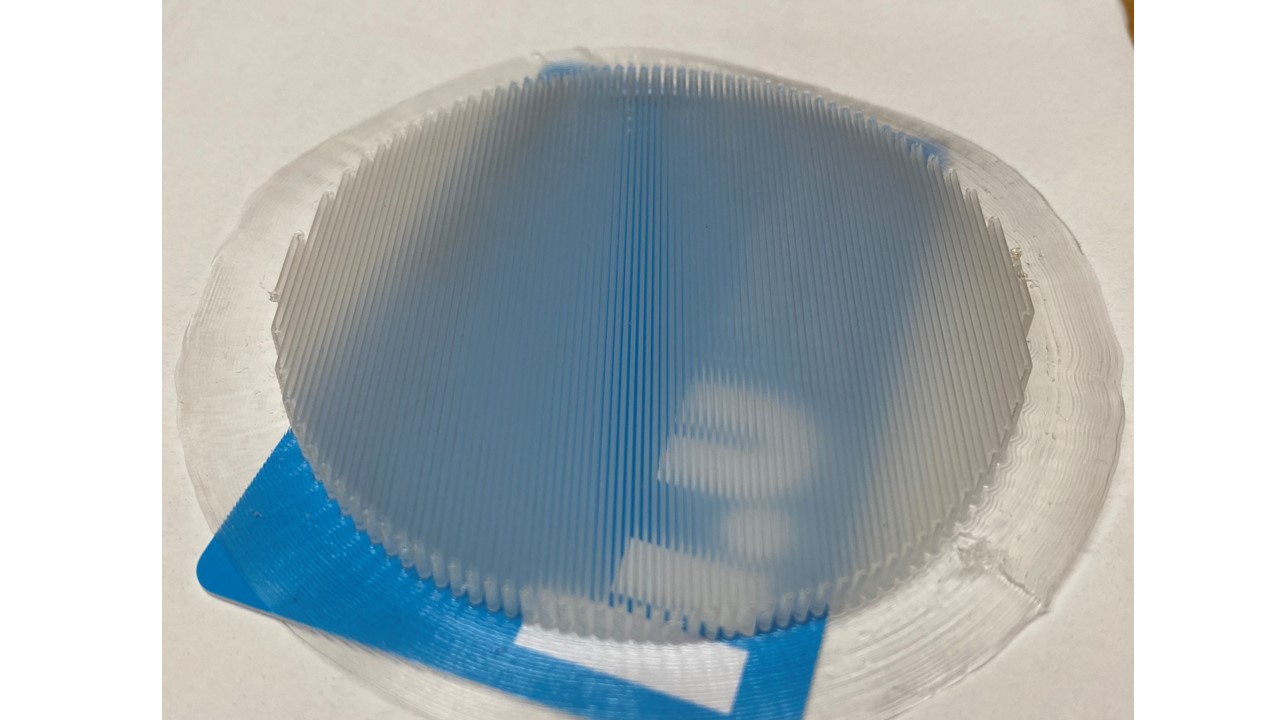}
    \caption{Terahertz meta wave plate (TMWP) used as anisotropic object which modulates the Stokes vector components of terahertz light transmitted through the wave plates upon rotation around the surface normal. The wave plates are obtained by 3D printing of columnar structures using printable polymers in sub mm sized column features.}
    \label{fig:TMWP}
\end{figure}

{To modulate all elements of the Stokes vector (see Eq.~\eqref{eq:Svector} below) of the incident terahertz radiation and to adequately demodulate the elements of the Stokes vector of the terahertz radiation emerging from the sample, we incorporated two 3D printed wave plates. The concept of 3D printing wave plates {is described in the work} by Hernandez-Serrano \textit{et al.},\cite{Hernandez} and has since undergone considerable development.\cite{perspective} For our setup, we adopted the design proposed by Rohrbach \textit{et al.}\cite{Rohrbach} The resulting structure, which forms the core of the TMWP, is shown in Fig.~\ref{fig:TMWP}. The feature sizes are sufficiently small to minimize scattering within the spectral range of interest, while the design maintains strong anisotropic properties, including both dichroism and birefringence. The structure consists of metagratings that are highly ordered and arranged on top of a support layer with {450}~$\mu$m thickness printed first. We note TWMPs with differing thicknesses may be needed for different spectral regions. The structures used here are sufficient to modulate the Stokes vector elements for the spectral range reported (170-250~GHz). The TMWPs are mounted into belt driven, ball bearing supported, non magnetic high precision rotation stages. The belt is connected to a stepper motor that controls the rotation of the TMWP. The motor is located remotely from the magnet to avoid magnetic field interference with the stepper motor operation. The motor is controlled by the electronic control board. Details of the rotation stage are shown in Fig~\ref{fig:detector_side}.}

\subsection{Superconducting Helmholtz magnet}
Our terahertz ellipsometer optical assembly at the Terahertz Materials Analysis Center incorporates a splitcoil superconducting magnet (Cryogenics Ltd., London, UK) capable of generating magnetic fields from -8~T to 8~T, with a field inhomogeneity of approximately 3000 parts per million across a central cylindrical volume of {10}~mm diameter {and length}. This magnet is equipped with three optical access windows, enabling transmission under a small range around near normal to surface incident angles and reflection measurements at 45$^{\circ}$ angle of incidence. The magnet was previously described for and used with an optical Hall effect instrument and further details are given in Refs.~\onlinecite{10.1063/5.0082353,KnightRSI2020, PhysRevB.110.054413, RichterPRBFebGO2024, rindert2024prl, RindertAPLCrbGO2024}. The instrument presented in this paper is thus an extension of this previous instrument, as shown in Fig.~\ref{fig:OHE+EPR}, where the optical Hall effect instrument can be seen to the right. The magnet is mounted on a rail system which permits to locate the magnet into both instruments for operation. The magnet further permits to operate as a sample cryostat, and the sample temperature can be adjusted between 4~K and {400~K}. The sample is held in a vacuum or low pressure helium atmosphere when needed for temperature control. {The magnet introduces non idealities and standing waves, that are not accounted for in the calibration. Generally, we address this by looking at difference data between two measurements performed with different magnetic field conditions, as explained in Sec.~\ref{SIC}.}

\subsection{Terahertz optical path}
Figure~\ref{fig:instrument} depicts the optical path of the terahertz ellipsometer with magnet. The terahertz beam emerges from the source D after passing through the frequency doubler (not shown). {The source is angled at 45$^\circ$ to maximize transmission through the first wire grid polarizer.} The terahertz beam is then collimated by gold coated off axis parabolic mirror M1 and transmitted through wire grid polarizer P45. The linear polarizer is rotated by azimuth angle  $\phi=45^{\circ}$ with respect to the plane of incidence and counter clockwise looking into the beam {towards the source}. Note that at $\phi=0^{\circ}$ or $90^{\circ}$ the electric field direction {would be} parallel or perpendicular to the plane of incidence, respectively. The beam then passes through the first TMWP. After the TMWP the beam is redirected by a gold coated, 90$^{\circ}$ off axis parabolic mirror M2 onto the sample. The terahertz beam thereby passes through a pair of outer and inner windows of the magnetocryostat which houses the sample. After passing the sample, the terahertz beam is collected either in transmission configuration (not shown) or in reflection configuration (as shown). In either case, the beam {propagates} through a second pair of windows, an inner window and an outer window exiting the magnetocryostat, and is collected by another gold coated, 90$^{\circ}$ off axis parabolic mirror M3 just outside the magnetocryostat. The focal lengths of the two off axis parabolic {mirrors} M2 and M3 ($f=210$~mm) are identical. The collimated terahertz beam is then redirected by plane mirror M4. The collimated beam then passes through another TMWP and linear polarizer P-45. The linear polarizer is rotated by azimuth angle  $\phi=-45^{\circ}$. Then another gold coated, 90$^{\circ}$ off axis parabolic mirror M5 focuses the terahertz beam onto quasi optical detector D. The detector unit of the instrument carrying the PSD (consisting of M3, M4, TWMP, Pol-45, M5) and D is mounted on a fixture, which itself is mounted on the arm of a $\theta-2\theta$ goniometer (not shown). The goniometer axis is perpendicular to the plane {depicted} in Fig.~\ref{fig:instrument}, and centered onto the sample surface (not shown). Figure~\ref{fig:detector_side} shows details of the detector unit. In transmission configuration, the goniometer mounted detector unit is moved into a straight line configuration to meet the direction of the incident beam. The optical path through the PSD mirrors the source side except that the wire grid polarizer is azimuthally rotated by -45$^{\circ}$ degrees. After passing through the PSD the terahertz radiation is detected by the quasi optical detector. To avoid multiple interference between optical components within the beam path, the linear polarizers are mounted with their surface at a small angle relative to the beam propagation direction.

\subsection{Electronic control board}
{The primary function of the control board is to precisely synchronize the two stepper motors driving the TMWPs within the PSG and PSD. Each TMWP is equipped with an encoder that sends signals to the control board. The measurement process begins when the sensor detects the encoder reference position, which mark the start of a cycle. After a full revolution, the encoder generates a home signal, which ensures proper alignment of the TMWPs and triggering data capture at the correct rotational positions. A PID control loop processes the encoder signals in real time to maintain precise positioning and synchronization. The motor speeds, set by the host computer, are regulated using the PID control algorithm to keep the TMWPs synchronized with data acquisition. The control board interfaces with a digital oscilloscope (Picoscope), which ensures precise coordination between TMWP motion, frequency sweep, and data collection for accurate and repeatable measurements.} 

\subsection{Instrument operation}
\begin{figure*}
    \centering
    \includegraphics[width=1\linewidth]{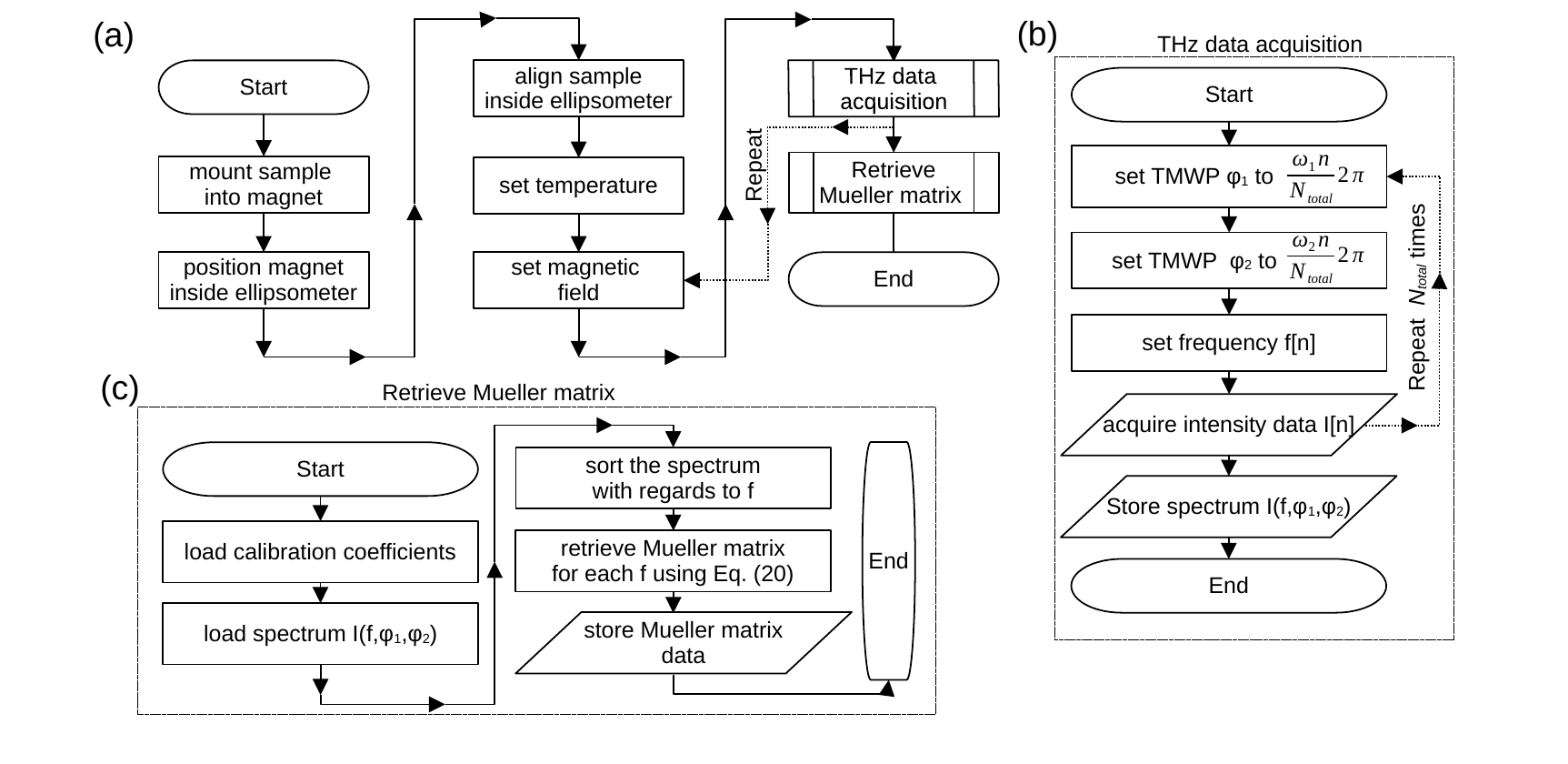}
    \caption{Flow chart of an HFEPR-GSE measurement (a), terahertz frequency scanning data acquisition procedure (b), and Mueller matrix data reduction from measurement data procedure (c).}
    \label{fig:Flowchart}
\end{figure*}

{The instrument operation is determined by the detector readout rate, the choice of sweep time, the number of sweeps, and the swept frequency interval. The control board provides signals to the Picoscope to indicate the start and end of each measurement frame. During this interval, the Picoscope collects the intensity data registered by the QOD. The Picoscope records the detector signal at a read out rate of 4~~$\mu$s. After the end of the measurement the collected intensity data set is then saved as one data frame to the host computer for subsequent analysis. The rotation speeds for PSG and PSD must be set. The choice should be such that both move at constant rotation but different speeds and such that both reach their original positions at the end of the measurement, i.e., after integer amounts of full rotations. During operation, in one example, the instrument continuously sweeps the source frequency over 80~GHz  frequency interval from 170 up to 250~GHz during a time interval of 150~ms, then back down to 170~GHz in another 150~ms, repeating this cycle. During each frequency sweep the picoscope continuously acquires intensity readings from the QOD. During the read out time of 4~$\mu$s during which one intensity data point is determined, in this example, the frequency moves up or down by 2.13~MHz, setting the frequency resolution. If 512 up and down sweeps are performed, in this example, each measurement requires a total time of 161.28~s. The synthesizer generates a trigger output signal that indicates the start and end of each up and down sweep. This trigger signal serves as a reference signal for the motion controller, ensuring synchronization between the frequency sweep and proper rotation of the PSG and PSD. During one full measurement cycle of 512 up and down sweeps, in this example, the speed of the PSG is set to complete one full revolution in 161.28~s while the PSD will complete seven full revolutions. This corresponds to a relative speed ratio of 1:7 (PSG to PSD) in this example. The overall workflow of the instrument operation and subsequent data analysis is illustrated in Fig.~\ref{fig:Flowchart}. A data sorting algorithm is then used to create a table where each intensity is associated with the mean frequency during which the intensity was collected and the corresponding settings of the rotation variables of PSG and PSD. The mean frequency value is determined as the average frequency value during the sweep with the read out time of 4~$\mu$s. Further downsampling may be performed numerically across multiple frequency rows in this table to increase the frequency interval over which data are averaged.}

\section{Instrument calibration}
A critical factor to the development of this instrument into a useful terahertz ellipsometer is the proper execution of calibration procedures. The calibration process aims to determine the effect of the rotation parameter settings in both PSG and PSD onto their respective modifications of any given Stokes vector traversing PSG and PSD at every frequency of interest. The calibration can be done in many different ways. Here, we render the effect of the rotation parameters onto the Stokes vector modification by a Fourier expansion in system transfer Mueller matrix elements for PSG and PSD separately. The calibration scheme follows the same principles as described previously by Ruder~\textit{et al.}\cite{Ruder:20} for a dual rotating anisotropic mirror full Mueller matrix ellipsometer setup operating at visible wavelengths.  The calibration described here includes measurements of well known and well characterized samples. Such samples consist of, but are not limited to, straight through air, reflection off highly reflecting surfaces such as gold mirrors, reflection or transmission on semi transparent samples such as doped semiconductor substrates, and highly anisotropic but well defined samples such as linear polarizers. The measurements consist of large sets of intensity values obtained at all frequencies under considerations and various settings of PSG and PSD rotation angles (azimuth angles of the TMWPs). All measurements from multiple calibration samples are then analyzed simultaneously, with each set of measurements per sample using appropriate model functions to render the effect of the respective sample Mueller matrix onto the detected intensities while searching for the \textit{a priori} unknown coefficients of the Fourier expansions for the PSG and PSD. When the calibration constants are known, any measurement performed on an unknown sample can then be numerically transformed by the use of the calibration data into the Mueller matrix elements of the unknown sample. In order to demonstrate the correct performance of the instrument, such measurement, when performed on another well known and well characterized sample, should result in accurate and precise Mueller matrix elements specific to that sample. Here, as one example, a highly anisotropic but transparent substrate is used for which the anisotropic dielectric constants are known, and the thickness is determined by varying model calculated data until the model data matches with the measurement obtained with this instrument after calibration. The thickness obtained is then compared to the result from the mechanical measurement.

\subsection{Jones and Mueller calculus}
Mueller matrix ellipsometers produce results conveniently described by the framework of the Stokes-Mueller calculus. At the core of the Stokes-Mueller calculus lies the Stokes vector $S$, which comprises four real valued elements that encapsulate the sums and differences in experimentally accessible intensities between various polarization states. A Cartesian coordinate system is needed to express all elements in this calculus. Commonly in ellipsometry, the plane of incidence is used whereby the Stokes vector is characterized by intensities measured transverse to the propagation direction using a linear polarization filter set parallel ($p$) to the plane of incidence, perpendicular ($s$) to the plane of incidence, rotated counter clockwise (+)/clockwise(-) by $\pm45^\circ$ from the plane of incidence, and using left circularly polarized (LCP) or right circularly polarized (RCP) filters. The following representation is then used in this work
\begin{equation}
    S = \begin{pmatrix}
        I_s + I_p \\ I_p - I_s \\ I_{45} - I_{-45} \\ I_{\mathrm{RCP}} - I_{\mathrm{LCP}}
    \end{pmatrix}.
    \label{eq:Svector}
\end{equation}
The Stokes vector undergoes transformation upon reflection or transmission via an optical component. This transformation is then described by the real valued Mueller matrix $M$, a $4\times4$ matrix of the form
\begin{equation}
    \textbf{M} = \begin{pmatrix}
        M_{11} & M_{12} & M_{13} & M_{14} \\
        M_{21} & M_{22} & M_{23} & M_{24} \\
        M_{31} & M_{32} & M_{33} & M_{34} \\
        M_{41} & M_{42} & M_{43} & M_{44} 
    \end{pmatrix}.
\end{equation}
\subsection{Mueller Matrix measurement}
The Stokes vector $S_{\mathrm{det}}$, representing the light incident on the detector, can be expressed as
\begin{equation}
    S_{\mathrm{det}} = M_{\mathrm{PSA}} M_{\mathrm{sam}} M_{\mathrm{PSG}} S_{\mathrm{sou}},
\end{equation}
where $M_{\mathrm{PSA}}$, $M_{\mathrm{sam}}$, and $M_{\mathrm{PSG}}$ denote the Mueller matrices corresponding to the analyzer, sample, and polarization state generator (PSG), respectively, and $S_{\mathrm{sou}}$ represents the Stokes vector of the light emitted from the source. The light incident on the sample can be represented as
\begin{equation}
    \hat{G}=M_{\mathrm{PSG}} S_{\mathrm{sou}}= M_{\mathrm{TMWP}}M_{\mathrm{45}} S_{\mathrm{sou}},
\end{equation}
where $M_{\mathrm{TMWP}}$ is the Mueller matrix of a TMWP and $M_{\mathrm{45}}$ is the Mueller matrix of a polarizer rotated by 45 degrees. {Here we assume an ideal source vector, $S_{\mathrm{sou}}=(1,0,0,0)$.} This expression can be further simplified to
\begin{equation}
    \hat{G}=M_{\mathrm{TMWP}}(1,0,1,0)^T\equiv(g_1,g_2,g_3,g_4)^T,
\end{equation}
where $T$ denotes the transpose. Previous studies on ellipsometer instruments which used the same operation concept\cite{Ruder:20, Ruder:21} have demonstrated that $\hat{G}$ can be effectively represented using a Fourier series with a sufficient number of coefficients. In this study, we employ a fourth order Fourier series, expanded over the azimuthal angle $\phi_1$ of the TMWP within the PSG
\begin{align}
    \begin{split}
        &\hat{G}(f,\phi_1) = (g_j)(f,\phi_1)=\\ 
        &a_{0,j} + \sum_{k=1}^4 \left( a_{k,j} \cos(\phi_1 k) - b_{k,j} \sin(\phi_1 k) \right).
    \end{split}
\end{align}
The polarization modulation of the generator is thus characterized by 36 Fourier coefficients, which are determined during the calibration process, as described in the subsequent sections. Since the detector is insensitive to the polarization state of the light, the experimentally measured quantity is the intensity $I_{\mathrm{det}}$. The intensity at the detector can be written as
\begin{align}
    \begin{split}
        &I_{\mathrm{det}} = (1, 0, 0, 0)S_{\mathrm{det}} = \\ 
        &(1, 0, 0, 0)M_{\mathrm{PSA}}M_{\mathrm{sam}}\hat{G} = \\
        & (d_1, d_2, d_3,d_4)M_{\mathrm{sam}}\hat{G} \equiv \hat{D}M_{\mathrm{sam}}\hat{G},
    \end{split}
\end{align}
where $\hat{D}$ is analogous to $\hat{G}$ and is similarly expressed as an {fourth} order Fourier series
\begin{align}
    \begin{split}
        &\hat{D}(f,\phi_2) = (d_j)(f,\phi_2)=\\ 
        &a_{0,j} + \sum_{k=1}^4 \left( a_{k,j} \cos(\phi_2 k) - b_{k,j} \sin(\phi_2 k) \right),
    \end{split}
\end{align}
where $\phi_2$ is the azimuthal angle of the TMWP on the detector side. The total number of calibration parameters is given by {$2\times36$}$\times N_{\mathrm{freq}}$, where $N_{\mathrm{freq}}$ is the total number of {mean frequency values}, which can be calculated as
\begin{equation}
    N_{\mathrm{freq}} = \frac{f_{\mathrm{end}} - f_{\mathrm{start}}}{\Delta f},
\end{equation}
where $f_{\mathrm{start}}$ and $f_{\mathrm{end}}$ denote the starting and ending frequencies of the interval, respectively, and $\Delta f$ represents the frequency step size.
{Because} the intensity is recorded at discrete time intervals $n=0,1,\ldots,N_{\mathrm{total}}-1$, the TMWP azimuthal angles are discretized as
\begin{align}
    &\phi_1[n] = \frac{\omega_1 n}{N_\mathrm{total}} 2\pi, \\
    &\phi_2[n] = \frac{\omega_2 n}{N_\mathrm{total}} 2\pi,
\end{align}
where $\omega_1$ and $\omega_2$ represent the number of revolutions per measurement cycle for the PSG and PSD TMWPs, respectively. The total number of points $N_\mathrm{total}$ is given by
\begin{equation}
    N_\mathrm{total} = 2 N_\mathrm{freq} N_\mathrm{sweeps},
\end{equation}
where $N_\mathrm{sweeps}$ is the number of sweeps per measurement cycle, with the factor 2 accounting for the intensity data collected during both the increasing and decreasing frequency sweeps. Similarly, the frequencies get discretized as
\begin{align}
    \begin{split}
        &f[n] = 2\left(\Delta f( N_\mathrm{freq} - 1)\right) \times\\ &\left|\frac{n + 1/2}{2N_\mathrm{freq}} -\left\lfloor\frac{n + 1/2}{2N_\mathrm{freq}} + \frac{1}{2} \right\rfloor \right| + f_\mathrm{start}
    \end{split}.
\end{align}
Here, $\lfloor\rfloor$ refers to the floor function. The intensity recorded by the detector at discrete time points during a full measurement is given by
\begin{equation}
    I_\mathrm{det}[n] =  \hat{D}^\mathrm{T}(f[n], \phi_2[n])M_\mathrm{sam}(f[n])\hat{G}(f[n], \phi_1[n]).
\end{equation}
To analyze the intensity as a function of frequency the values must be sorted accordingly. The intensity recorded by the detector for a single frequency can then be expressed as
\begin{align}
\begin{split}
&I_\mathrm{det}[n_\nu] = \\& \hat{D}^\mathrm{T}(f[n_\nu], \phi_2[n_\nu])M_\mathrm{sam}(f[n_\nu])\hat{G}(f[n_\nu], \phi_1[n_\nu]) \\
&= \sum_{i,j}^4 d_i(f[n_\nu], \phi_2[n_\nu])M_{ij}(f[n_\nu])g_j(f[n_\nu], \phi_1[n_\nu]).
\end{split}
\label{Eq:c1}
\end{align}
where the subset of indices $n_\nu$ corresponding to a single frequency $\nu$ is given by
\begin{equation}
    n_\nu = k, 2N_\mathrm{freq} \pm k, 4N_\mathrm{freq} \pm k, \dots, N_\mathrm{total} - k - 1,
    \label{Eq:c3}
\end{equation}
and where
\begin{equation}
    k[f] = N_\mathrm{freq} \frac{f - f_\mathrm{start}}{f_\mathrm{start}}.
\end{equation}
The resulting ($2\times N_\mathrm{sweep}$)-dimensional intensity vector $I_\mathrm{det}$ is then defined as
\begin{equation}
    I_\mathrm{det} = A x_\mathrm{sam,\nu},
    \label{eq:int}
\end{equation}
for a given frequency, where $A$ is an $(2\times N_\mathrm{sweep})\times16$ matrix containing the elements of $\hat{G}$ and $\hat{D}$, defined as
\begin{align}
    \begin{split}
        &A_\nu = \begin{pmatrix}
            (d_1g_1)(n=0) & (d_1g_2)(n=0) & \dots & (d_4g_4)(n=0) \\
            (d_1g_1)(n=1) & (d_1g_2)(n=1) & \dots & (d_4g_4)(n=1) \\
            \vdots & \vdots & \ddots & \vdots \\
            (d_1g_1)(n=N) & (d_1g_2)(n=N) & \dots & (d_4g_4)(n=N)
        \end{pmatrix}.
    \end{split}
    \label{Eq:c4}
\end{align}
Here, $x_\mathrm{sam,\nu} = (M_{11}(f), M_{12}(f), \dots, M_{44}(f))^\mathrm{T}$ is a 16 element vector. In principle, acquiring at least eight frequency sweeps ($N_\mathrm{sweeps} > 8$) is sufficient to determine the sample Mueller matrix elements. 

Finally, a pseudoinverse is applied to $A_\nu$ to retrieve the Mueller matrix of the sample for the frequency $\nu$, i.e., measurement of the full $4\times4$ Mueller matrix
\begin{equation}
    x_\mathrm{sam,\nu} = (A_\nu^T A_\nu)^{-1} A_\nu^T I_\mathrm{det} = A_\mathrm{red} I_\mathrm{det}.
    \label{Eq:c5}
\end{equation}
The steps outlined in Eqs.~\eqref{Eq:c1}-\eqref{Eq:c5} are repeated for each frequency.

\subsection{Calibration samples}
A series of well characterized reference samples, including isotropic and anisotropic materials, were used for instrument calibration. 

\subsubsection{{Straight through air}}
The calibration procedure included measurements in straight through air configuration, i.e., with the magnet removed and the instrument set to transmission configuration. Then, the intensities are recorded without any sample present except for normal ambient. To model the effect of the normal ambient, the identity Mueller matrix is used
\begin{equation}
    M_\mathrm{Air}= \begin{pmatrix}
        1 & 0 & 0 & 0 \\
        0 & 1 & 0 & 0 \\
        0 & 0 & 1 & 0 \\
        0 & 0 & 0 & 1 
    \end{pmatrix}.
\end{equation}

\subsubsection{Linear polarizer}
A wire grid linear polarizer is included in normal transmission configuration with the polarizer placed into the position of the sample. Measurements are repeated with the orientation of the linear polarizer set to specific azimuthal angles ($\phi=0^{\circ}$, $\phi=45^{\circ}$, $\phi=90^{\circ}$, $\phi=135^{\circ}$). To model the effect of the linear polarizer, the Mueller matrix of an ideal polarizer is used\cite{Fujiwara,TOMPKINS2005xv}

\begin{align}
\begin{split}
    &M_\mathrm{Pol}(\hat{\phi}) = \\ &\begin{pmatrix}
        1 & \cos(2\hat{\phi}) & \sin(2\hat{\phi}) & 0 \\
        \cos(2\hat{\phi}) & \cos^2(2\hat{\phi}) & \cos(2\hat{\phi})\sin(2\hat{\phi}) & 0 \\
        \sin(2\hat{\phi}) & \sin(2\hat{\phi})\cos(2\hat{\phi}) & \sin^2(2\hat{\phi}) & 0 \\
        0 & 0 & 0 & 0 \\
    \end{pmatrix}.
    \end{split}
\end{align}

where $\hat{\phi}=\phi-\phi_0$, and $\phi_0$ is the azimuth offset of the linear polarizer. During the calibration analysis, the specific azimuth angles $\phi$ are used as input parameters{ together with an azimuthal offset parameter for each measurement, which are then} searched for as calibration parameters. 

\subsubsection{$n$ type silicon substrate}
A $n$ type doped silicon wafer with low doping concentration is used in reflection configuration at multiple angles of incidence. In particular, for this calibration measurement the assembly incorporation of the $\theta-2\theta$ goniometer is critical. The main information gained from this measurement is to identify and differentiate the meaning of $p$ versus $s$ directions relative to the plane of incidence. It is noted that the latter is solely defined by the axis of the double goniometer. Measurements were collected at angles of incidence $\Phi_a =45^{\circ}$, $60^{\circ}$, and $75^{\circ}$. To model the effect of the silicon substrate, the following generic Mueller matrix in the so called ``$NSC$''-configuration was used\cite{Fujiwara,TOMPKINS2005xv}
\begin{equation}
    M_\mathrm{Si} = \begin{pmatrix}
        1 & -N & 0 & 0 \\
        -N & 1 & 0 & 0 \\
        0 & 0 & C & S \\
        0 & 0 & -S& C
    \end{pmatrix}.
\end{equation}
Here, $N$, $S$, and $C$ are defined as
\begin{align}
    &N = \cos(2\Psi), \\
    &S = \sin(2\Psi)\sin(2\Delta),\\
    &C = \sin(2\Psi)\cos(2\Delta),
\end{align}
where $\Psi$ and $\Delta$ are the ellipsometric angles. These angles differ for every frequency and every setting of $\Phi_a$ and were considered parameters to be determined during the optimization procedure.

\subsection{Calibration example}
\begin{figure*}
    \centering
    \includegraphics[width=1\linewidth]{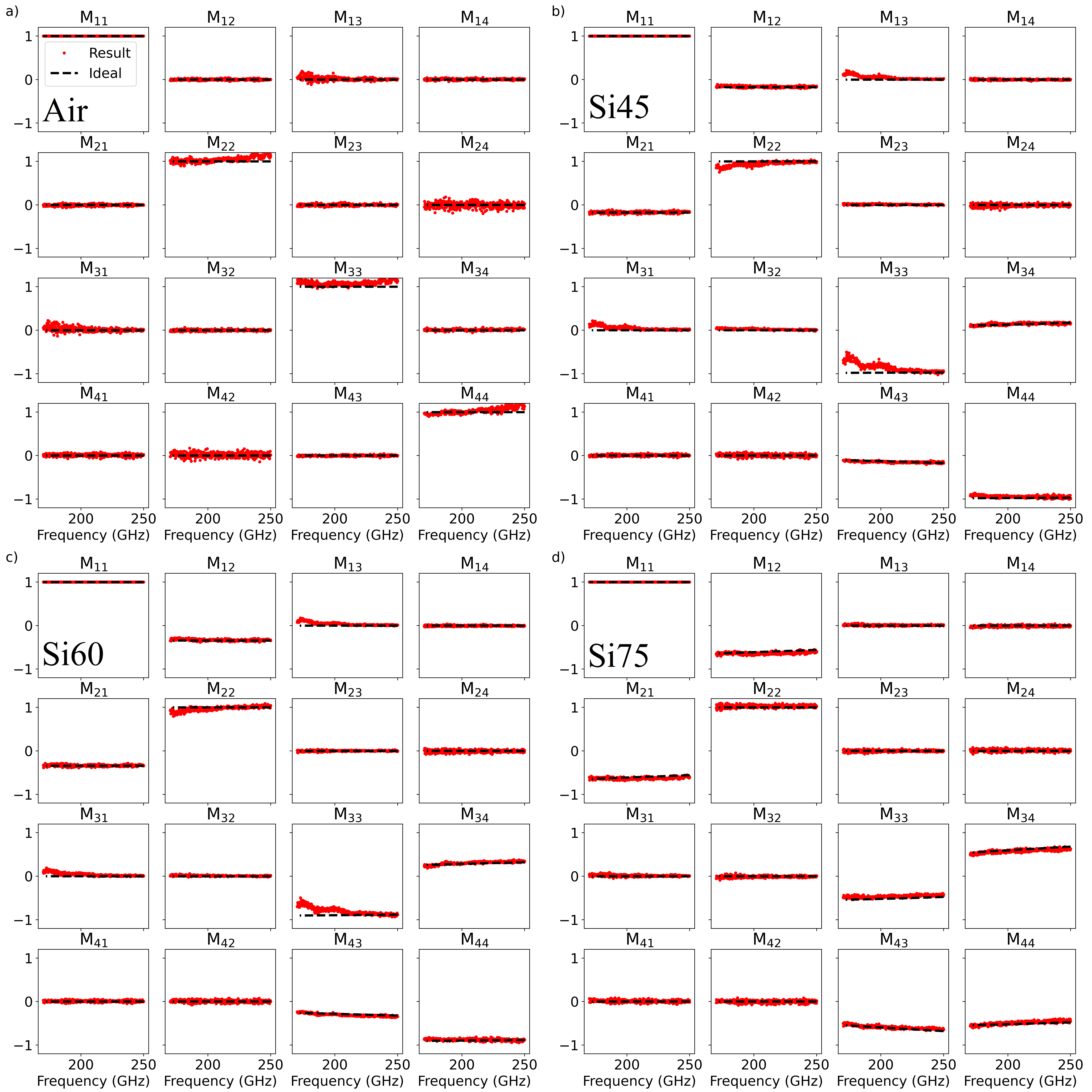}
    \caption{{Mueller matrices obtained from measurements of the $n$ type silicon substrate and straight through air for instrument calibration. Dashed black lines represent ideal model calculated Mueller matrix spectra. All other spectra are obtained from the calculations, searching for the best match calibration constants while minimizing the difference between the measured and calculated intensities. For the calculated intensities, the model Mueller matrix is the input, and calibration parameters are varied until intensities match. Then, the calibration constants are used to invert for a Mueller matrix, which is then shown here and thought of as  "measured". The numerical value of the labels following "Si" denotes the angle of incidence. The calibration was performed over a frequency range of 170-250~GHz in 411 steps and using 512 frequency sweeps per measurement. See text for further details.}}
    \label{fig:sis}
\end{figure*}
\begin{figure*}
    \centering
    \includegraphics[width=1\linewidth]{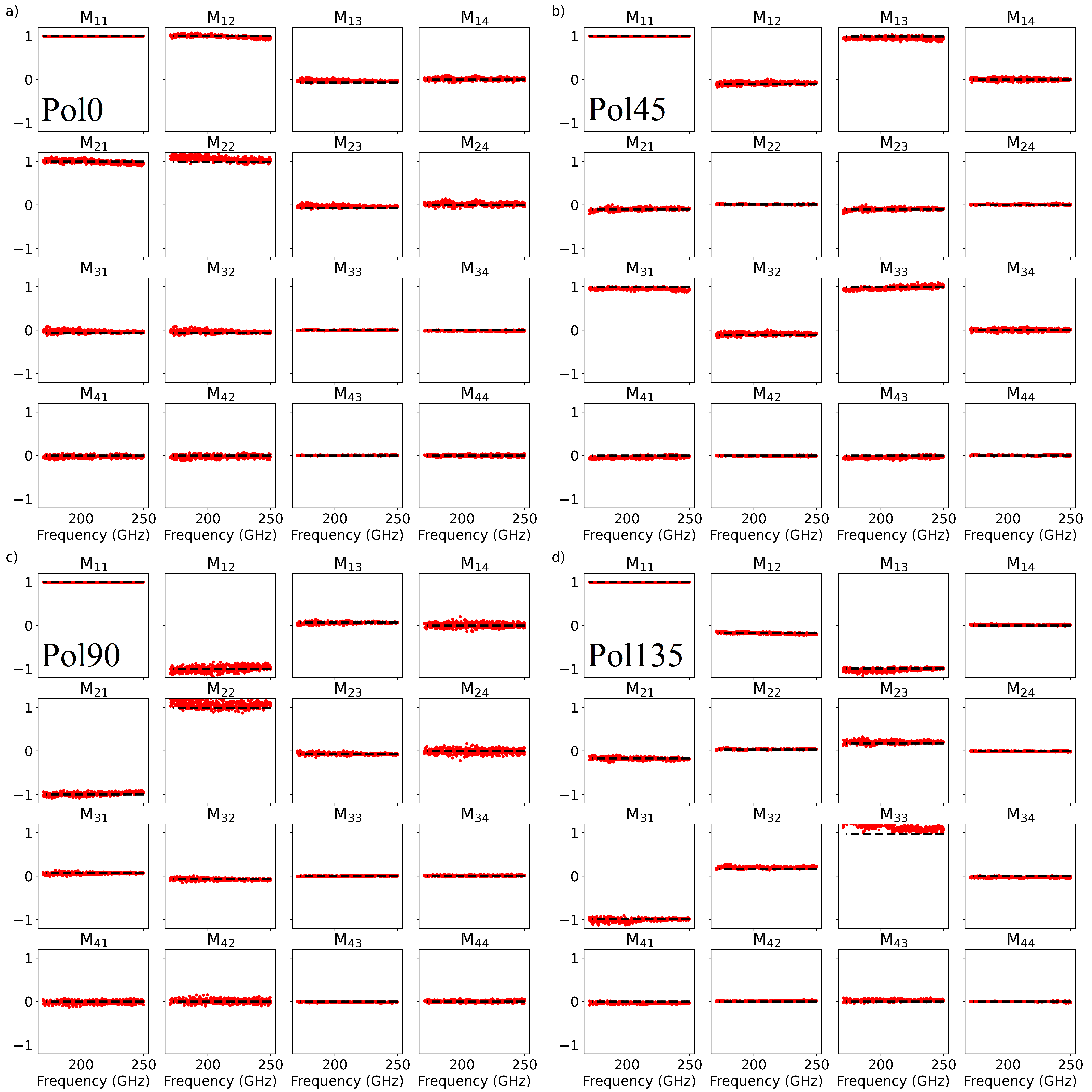}
    \caption{Same as Fig.~\ref{fig:sis} for the linear polarizer used in the instrument calibration. The numerical after "Pol" denotes the azimuthal rotation setting $\phi$. See text for further details.}
    \label{fig:pols}
\end{figure*}
\begin{figure*}[!tbp]
    \centering
    \includegraphics[width=0.98\linewidth]{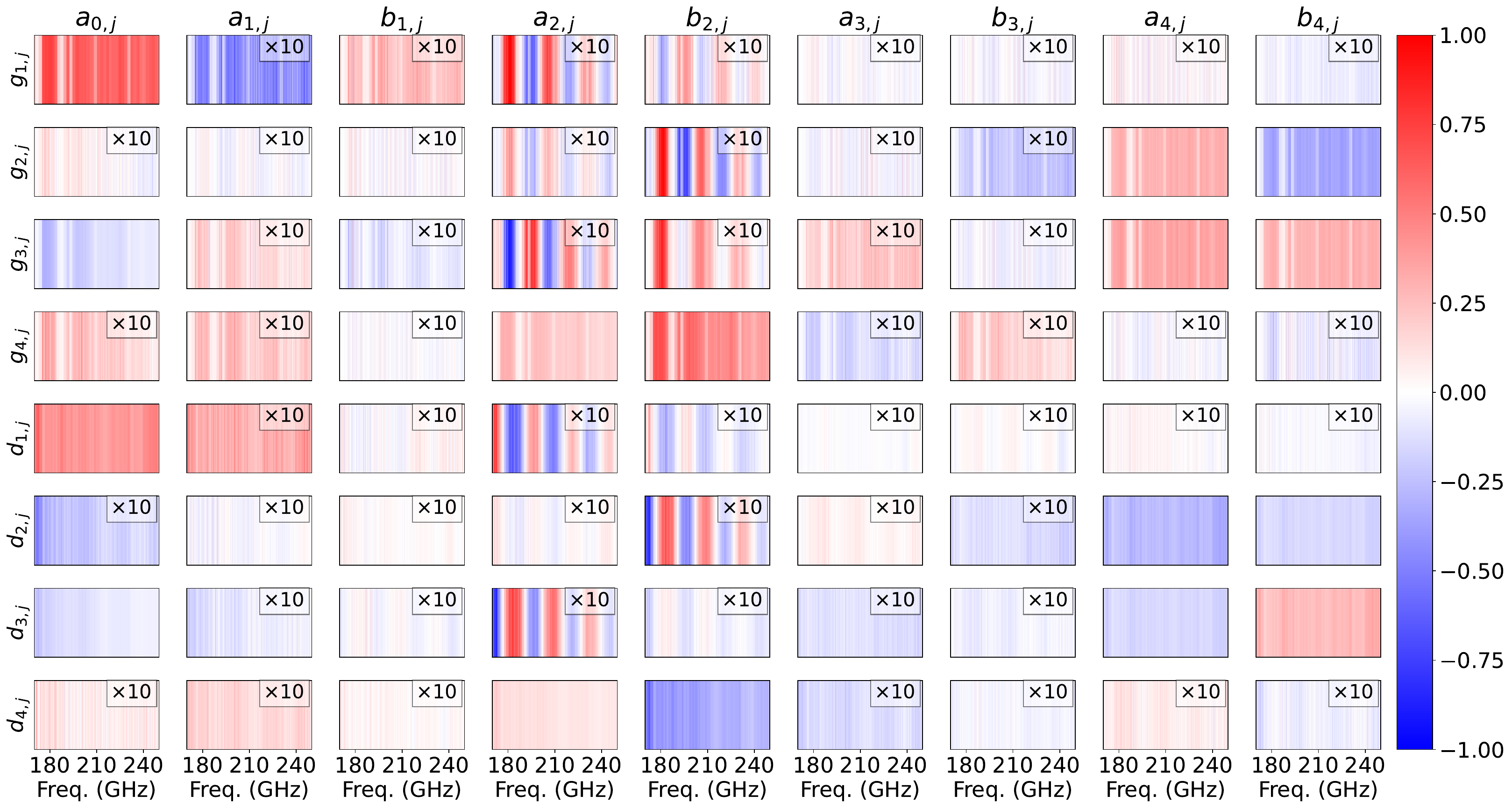}
    \caption{Stripe diagrams are shown for improved visibility for all 72 Fourier coefficients as a function of frequency, within the interval of 170-250~GHz, as an example and determined by the calibration process. Some elements, as indicated, are multiplied by 10 to enhance their visibility.}
    \label{fig:coeffs}
\end{figure*}
Calibration measurements were conducted over a frequency range from $f_\mathrm{start}=190$~GHz to $f_\mathrm{end}=230$~GHz, with $N_\mathrm{steps}=411$ frequency steps and $N_\mathrm{sweeps}=512$ sweeps per measurement. {The rotation speeds were set to reach one full revolution for PSG and seven full revolution for PSD per full measurement cycle.} Each measurement was repeated five times, and the results were averaged to improve accuracy. Calibration coefficients were determined by implementing a non linear least squares optimization algorithm that minimizes the difference between the measured and calculated intensities. The calculated intensity was derived using Eq.~\eqref{eq:int}, with $x_\mathrm{sam}$ calculated by the respective ideal Mueller matrix elements of the corresponding calibration sample model discussed above. In this process, the calibration parameters are varied, and the model Mueller matrix parameters are also varied where applicable, such as the azimuthal offset for the linear polarizer. Then, the calibration constants obtained from the best match for the calculated intensities are used to calculate the Mueller matrix {observed} during each calibration sample measurement. These Mueller matrix elements are shown in Figs.~\ref{fig:sis} and \ref{fig:pols} as ``measured'' data, {to assess the consistency of the obtained calibration parameters}. The strong agreement between the ``measured'' and model calculated Mueller matrices confirms the effectiveness of the calibration procedure and demonstrates that the continuously rotating TMWPs provide sufficient modulation of the Stokes vector elements in the terahertz spectral range. To quantify the accuracy of the instrument, the average root mean square error (RMSE) between the measured and ideal values, \(\Delta M_{i,j}\), was calculated as
\begin{equation}
    \mathrm{RMSE} = \sqrt{\frac{1}{N} \sum \Delta M_{i,j}^2} = 0.048,
\end{equation}
where the sum is taken over all calibration measurements, resulting in \(N=8\times16\times411=52,608\). In Fig.~\ref{fig:coeffs}, the 72 Fourier coefficients are plotted as a function of frequency, with those of smaller magnitude scaled by a factor of ten for clarity. It is evident that most of the calibration coefficients are close to zero, although some exhibit clear oscillations due to non idealities in the optical components. These observations underscore the necessity of incorporating these calibration coefficients into the data analysis.

\section{Examples and instrument verification}
\subsection{Thickness determination of a transparent, anisotropic substrate}
\begin{figure*}[!tbp]
    \centering
    \includegraphics[width=0.92\linewidth]{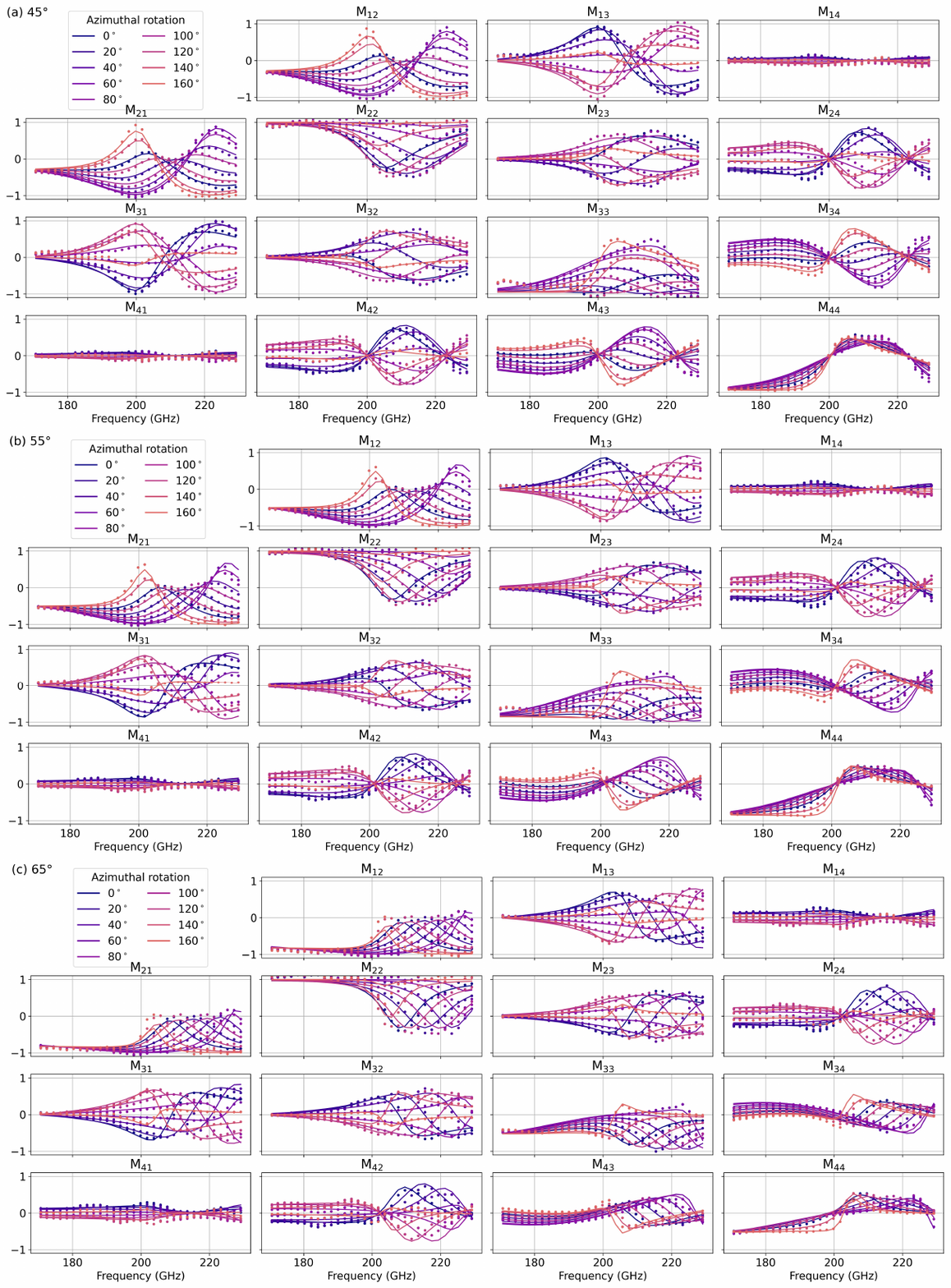}
    \caption{Comparison between measured and best match optical model calculated normalized Mueller matrix spectra for an $m$-plane sapphire substrate. The measurements were conducted at incident angles of (a) 45$^{\circ}$, (b) $55^{\circ}$, and (c) $65^{\circ}$, with in plane sample rotations at 20$^{\circ}$ intervals. The optical model consists of a single optically uniaxial layer with assumed dielectric constants of $\epsilon_\perp=9.389$ and $\epsilon_\parallel=11.614$, oriented with its optical axis parallel to the surface of the substrate, was optimized for the sample thickness, determined to be 0.45~mm. The good agreement between the measured and modeled data demonstrates the successful operation of the instrument as a full terahertz $4\times4$ Mueller matrix spectroscopic ellipsometer.}
    \label{fig:enter-label}
\end{figure*}
\begin{figure*}
    \centering
    \includegraphics[width=0.8\linewidth]{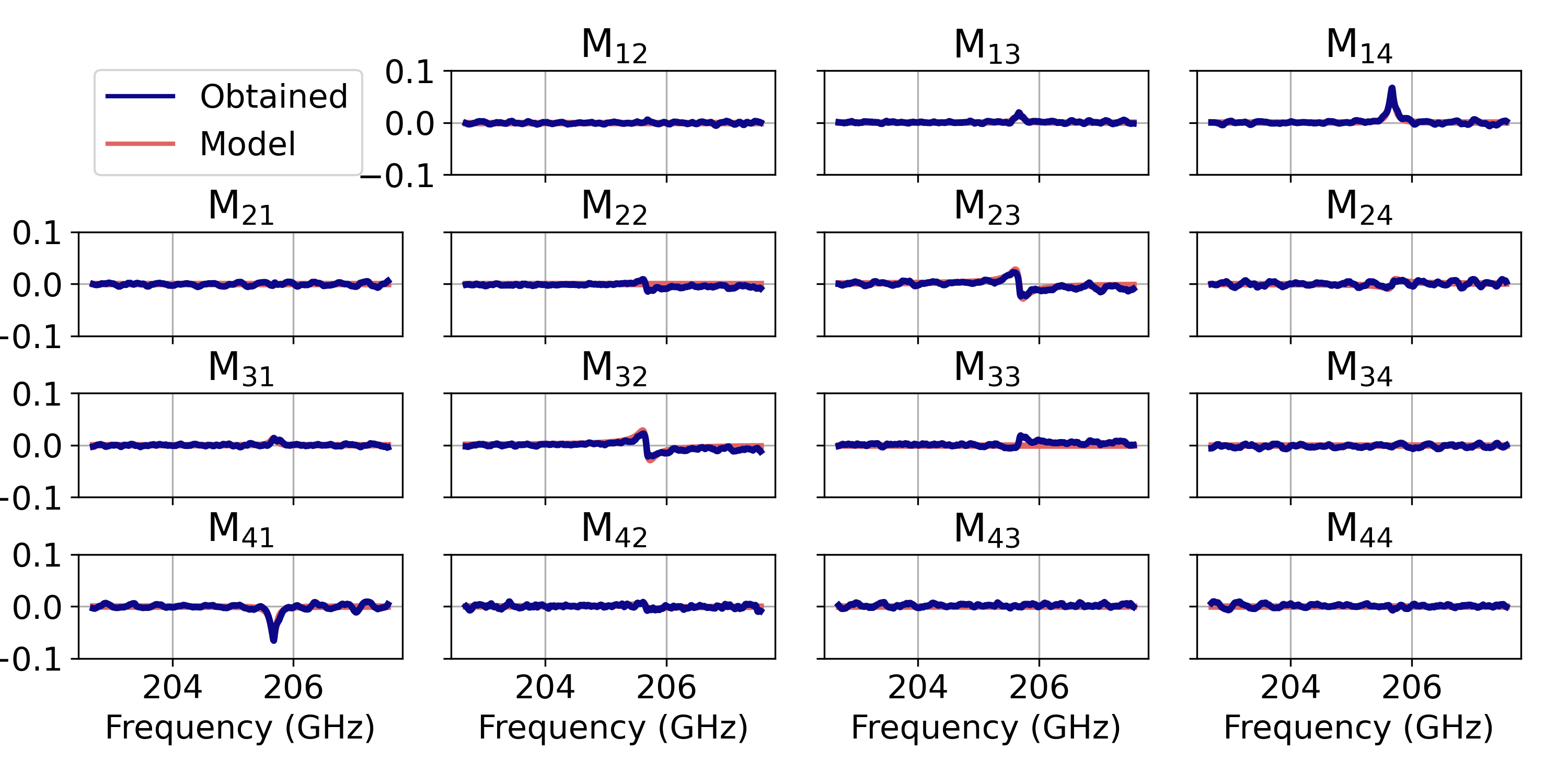}
    \caption{Difference Mueller matrix data for nitrogen doped 4H-SiC measured at a sample temperature of 10 K, with applied magnetic fields of $B = \pm 7.33$ T. The blue line represents the difference between the measurements taken at $B=7.33$ T and $B=-7.33$ T, effectively isolating the electron paramagnetic resonance signal. The orange lines correspond to the best match model calculated data using the Bloch-Brillouin formalism described previously by Rindert~\textit{et al.} in Ref.~\onlinecite{PhysRevB.110.054413}.}
    \label{fig:sic}
\end{figure*}
The calibration of the instrument was validated by measuring a commercially available, {2 inch} diameter, $m$-plane $(1\Bar{1}00)$ sapphire substrate. The sample was measured at incident angles of $\Phi_a=45^{\circ}$, $55^{\circ}$, and $65^{\circ}$. For each incident angle, measurements were taken at 20$^{\circ}$ intervals of in plane sample rotation, from 0$^{\circ}$ to 180$^{\circ}$ degrees. A total of three measurement cycles were accumulated for each rotational position of the sample. The collected data were subsequently processed to yield normalized Mueller matrix data. An optical model was then developed and calculated Mueller matrix spectra were best match to the measured data by varying model parameters. The optical model consisted of a single, free standing, optically uniaxial layer with thickness $d$, and with known dielectric constants, $\epsilon_\perp=9.389$ and $\epsilon_\parallel=11.614$, as reported in Ref.~\onlinecite{PhysRevB.61.8187} for the static anisotropic dielectric constants of sapphire. The $4\times4$ matrix algorithm\cite{PhysRevB.53.4265} was used to calculate the Mueller matrix elements, as summarized recently.\cite{PhysRevB.110.054413} Then, the only parameter optimized for was the model layer thickness $d$, determined to be 0.45~mm. The nominal thickness of the sapphire substrate is {0.44}~mm.

The Mueller matrix elements obtained from measurement using the instrument reported here and from the best match model analysis are presented in Fig.~\ref{fig:enter-label}. The agreement between the experimental and best match model is very good. Notably, the strong variation with frequency due to the multiple interference within the anisotropic {sample}, and its variation upon substrate rotation due to the variation of the orientation of the optic axis is captured by the measured data very closely. The results obtained here successfully demonstrate correct instrument performance. 

To improve the sensitivity, a metal plate could be placed under the sample to induce more reflections within the sample, thus helping with the decorrelation of the dielectric constants and the sample thickness. Possible pathways to further improve the agreement between the experimental and best match model calculated data are to include more reference samples during the calibration and to investigate and include non idealities of the optical components, such as the non linearity of the quasi optical detector. This will be investigated further in future work.

\begin{figure*}[!tbp]
    \centering
    \includegraphics[width=0.65\linewidth]{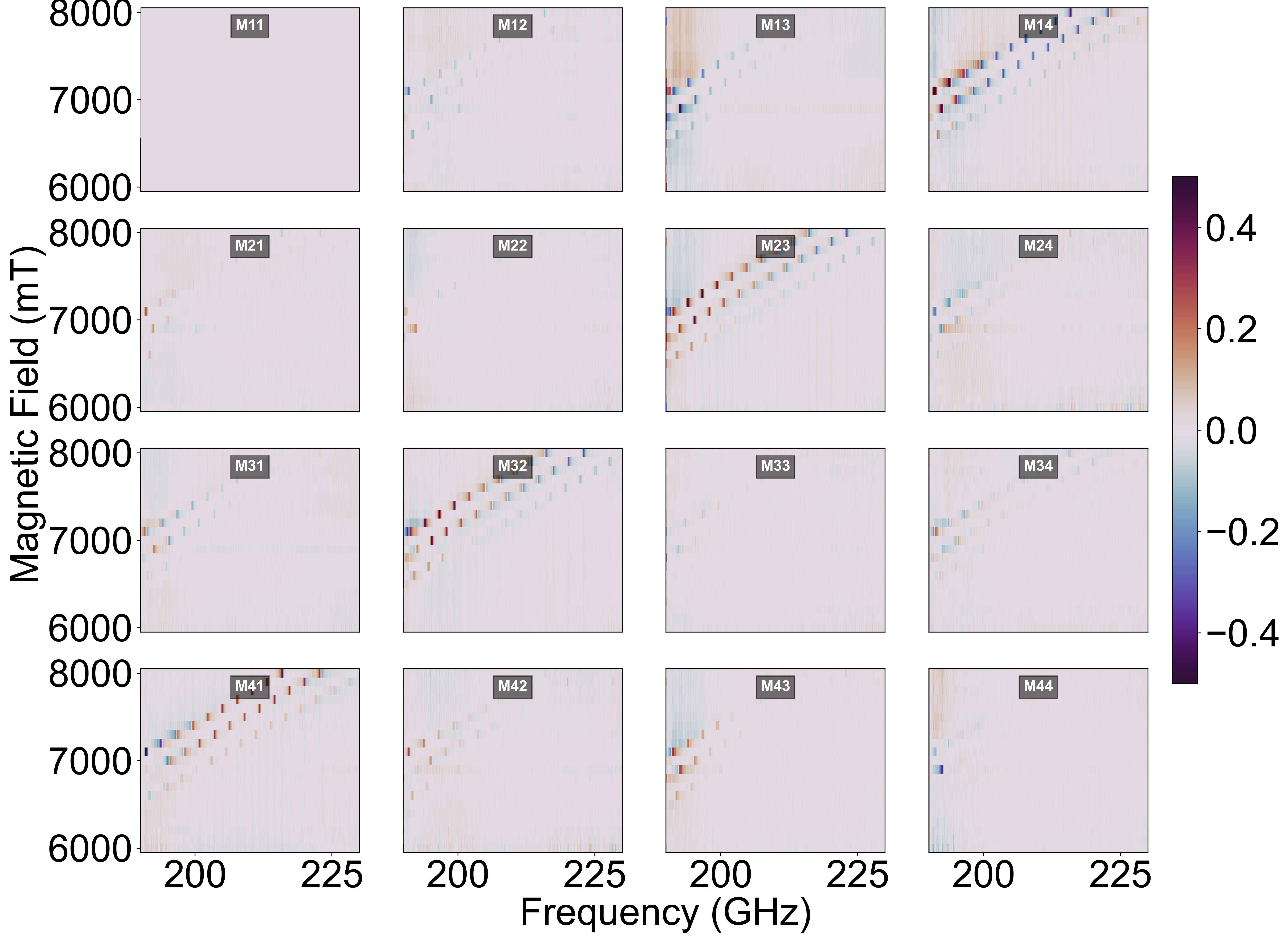}
    \caption{{Magnetic field vs. frequency map (a.k.a. Zeeman diagram) obtained from HFEPR-GSE measurements obtained at 15~K. The measurements were performed by sweeping the frequencies continuously between 190 and 230~GHz with the magnetic field fixed. This was repeated with the magnetic field from 6 to 8~T in steps of 0.1~T. The average of all measurements has been subtracted from each separate scan and Mueller matrix element to leave only the magnetic field dependent features visible (difference approach).}}
    \label{fig:Zeeman}
\end{figure*}
\subsection{Electron paramagnetic resonance due to the nitrogen defect in SiC:N\label{SIC}}
In this example, the superconducting magnet is placed into the optical assembly, and the frequency domain terahertz $4\times4$ Mueller matrix ellipsometer is demonstrated for use in magnetic resonance spectroscopy. Electron paramagnetic resonances can be detected by applying a magnetic field to a sample while performing terahertz ellipsometry. These resonances are typically small, with peak to peak variations in the Mueller matrix data on the order of a few 0.01 units.\cite{10.1063/5.0082353, RichterPRBFebGO2024,PhysRevB.110.054413,rindert2024prl} To enhance the detection of such subtle resonances, we employ a difference measurement technique, where differences are analyzed between measured spectra obtained with field applied and with no field applied. This approach offers several advantages, including a significant reduction of non idealities introduced by the superconducting splitcoil magnet and the elimination of magnetic field independent artifacts, such as {standing waves}, from the experimental data.\cite{10.1063/5.0082353, RichterPRBFebGO2024,PhysRevB.110.054413,rindert2024prl} The windows covering the variable temperature insert (VTI) inside the superconducting splitcoil magnet introduce standing waves and other non idealities not accounted for by the calibration procedure. However, these non idealities are also effectively canceled out through the difference measurement technique.

To illustrate this method, measurements were conducted on nitrogen doped 4H-SiC. Schubert~\textit{et al.}\cite{10.1063/5.0082353} reported recently on similar measurements using a terahertz ellipsometer setup capable of measuring the upper $3\times3$ block of the Mueller matrix only while employing rotating linear polarizers.\cite{2018LuEllipsometer} With the new instrument described here measurements were performed with the sample temperature held at 10~K, frequency swept from 202.7~GHz to 207~GHz in 500 steps, with magnetic field strengths of $B=\pm7.33$~T. The measurement at $B=-7.33$~T was subtracted from that at $B=7.33$~T to isolate the magnetic resonance and facilitate subsequent analysis. The resulting difference Mueller matrix data, along with best model calculated data are presented in Fig.~\ref{fig:sic}. The model used here is based on the Bloch-Brillouin formalism introduced in Ref.~\onlinecite{PhysRevB.110.054413}, where the permeability tensor $\boldsymbol{\upmu}(\omega)$ is expressed as

\begin{align}
    \begin{split}
    &\boldsymbol{\upmu}(\omega) = \boldsymbol{I} + \chi_{0} \begin{pmatrix}
        \frac{\omega_0^2}{\omega_0^2 - \omega^2 - i\omega\gamma}& \frac{i\omega\omega_0}{\omega_0^2 - \omega^2 - i\omega\gamma} & 0 \\
        \frac{-i\omega\omega_0}{\omega_0^2 - \omega^2 - i\omega\gamma} & \frac{\omega_0^2}{\omega_0^2 - \omega^2 - i\omega\gamma} & 0 \\
        0 & 0 & 1
    \end{pmatrix},
    \end{split}
    \label{eq:prbtensor_sum_broadened}
\end{align}
{for $B\parallel z$,} where $\gamma$ is the broadening parameter, $\omega_0=g\mu_BB$ is the resonance frequency, $\mu_B$ is the Bohr magneton, $g$ is the g-factor, and the dc susceptibility $\chi_0$ is modeled as
\begin{equation}
    \chi_0 = \frac{g\mu_B}{2B}\frac{1 - e^{-\hbar\omega_0/kT} } {1+e^{-\hbar\omega_0/kT}} n_s,
\end{equation}
where $k$ is the Boltzmann constant and $n_s$ is the concentration of unpaired electrons. As depicted in Fig.~\ref{fig:sic}, this model, with parameters $g=2.003$ and $n_s = 1.4\times 10^{17}$~cm$^{-3}$, provides an excellent match to the experimentally obtained data.
\subsection{Zeeman diagram of Cr doped $\beta$-Ga$_2$O$_3$}
To demonstrate the capability of the instrument to obtain Zeeman diagrams, measurements were performed on a Cr doped $\beta$-Ga$_2$O$_3$ sample with dimensions of 15$\times$15$\times$0.36~mm$^3$, cooled to 15~K. The Zeeman diagram was generated by conducting HFEPR-GSE measurements across a frequency range of 190 to 230~GHz at a fixed magnetic field, with subsequent measurements taken in 0.1~T steps from 6 to 8~T. A difference approach was employed, where the average of all measurements was subtracted from each individual measurement, effectively eliminating any features that are not dependent on the magnetic field. The result is shown in Fig.~\ref{fig:Zeeman}. Here, three visible magnetic resonances depend linearly on the magnetic field strength. Furthermore, the strength of the resonances is strongly frequency dependent. We propose that this is due to Fabry-Perot interference within the sample, effectively meaning that the sample itself acts like a cavity. This effect has been observed in previous HFEPR-GSE studies.\cite{10.1063/5.0082353} Analysis of the Zeeman diagram and a more detailed sample description are given elsewhere.\cite{RindertAPLCrbGO2024}

\section{Conclusion}
In this work, we successfully demonstrated the development and implementation of a terahertz 4$\times4$ Mueller matrix ellipsometer, capable of operating across a wide frequency range with high resolution. This instrument overcomes the limitations of previous terahertz ellipsometers by enabling Mueller matrix measurements and significantly improving data acquisition speed through fast frequency scanning and quasi optical detection systems, {ultimately enabling the acquisition of Zeeman diagrams.} The thickness of an anisotropic sapphire substrate could be determined with the help of an optical model, showing great agreement between experimental and calculated data. Furthermore, the instrument’s ability to accurately measure subtle magnetic resonance phenomena was validated through EPR measurements on nitrogen doped SiC. The difference measurement techniques effectively isolated magnetic resonance signals, confirming the sensitivity and versatility of the system. Future enhancements, such as refining calibration methodologies and incorporating additional calibration samples, promise to further extend the instrument’s capabilities. 

\section{Acknowledgments}
This work is supported by the Swedish Research Council under Grants No. 2016-00889 and No. 2022-04812, by the Knut and Alice Wallenberg Foundation funded grants “Wide-bandgap semiconductors for next generation quantum components” (Grant No. 2018.0071) and “Transforming ceramics into next generation semiconductors“ (Grant No.2024.0121), by the Swedish Governmental Agency for Innovation Systems VINNOVA under the Competence Center Program Grant No. 2022-03139, and by the Swedish Government Strategic Research Area NanoLund. M.S. acknowledges support by the National Science Foundation under awards ECCS 2329940, and OIA-2044049 Emergent Quantum Materials and Technologies (EQUATE), by Air Force Office of Scientific Research under awards FA9550-21-1-0259, and FA9550-23-1-0574 DEF, and by the University of Nebraska Foundation. M.S. also acknowledges support from the J.~A.~Woollam Foundation.

\bibliography{refs}

\begin{thebibliography}{63}%
\makeatletter
\providecommand \@ifxundefined [1]{%
 \@ifx{#1\undefined}
}%
\providecommand \@ifnum [1]{%
 \ifnum #1\expandafter \@firstoftwo
 \else \expandafter \@secondoftwo
 \fi
}%
\providecommand \@ifx [1]{%
 \ifx #1\expandafter \@firstoftwo
 \else \expandafter \@secondoftwo
 \fi
}%
\providecommand \natexlab [1]{#1}%
\providecommand \enquote  [1]{``#1''}%
\providecommand \bibnamefont  [1]{#1}%
\providecommand \bibfnamefont [1]{#1}%
\providecommand \citenamefont [1]{#1}%
\providecommand \href@noop [0]{\@secondoftwo}%
\providecommand \href [0]{\begingroup \@sanitize@url \@href}%
\providecommand \@href[1]{\@@startlink{#1}\@@href}%
\providecommand \@@href[1]{\endgroup#1\@@endlink}%
\providecommand \@sanitize@url [0]{\catcode `\\12\catcode `\$12\catcode `\&12\catcode `\#12\catcode `\^12\catcode `\_12\catcode `\%12\relax}%
\providecommand \@@startlink[1]{}%
\providecommand \@@endlink[0]{}%
\providecommand \url  [0]{\begingroup\@sanitize@url \@url }%
\providecommand \@url [1]{\endgroup\@href {#1}{\urlprefix }}%
\providecommand \urlprefix  [0]{URL }%
\providecommand \Eprint [0]{\href }%
\providecommand \doibase [0]{https://doi.org/}%
\providecommand \selectlanguage [0]{\@gobble}%
\providecommand \bibinfo  [0]{\@secondoftwo}%
\providecommand \bibfield  [0]{\@secondoftwo}%
\providecommand \translation [1]{[#1]}%
\providecommand \BibitemOpen [0]{}%
\providecommand \bibitemStop [0]{}%
\providecommand \bibitemNoStop [0]{.\EOS\space}%
\providecommand \EOS [0]{\spacefactor3000\relax}%
\providecommand \BibitemShut  [1]{\csname bibitem#1\endcsname}%
\let\auto@bib@innerbib\@empty
\bibitem [{(2007)}]{Fujiwara}%
  \BibitemOpen
  \enquote {\bibinfo {title} {Principles of optics},}\ in\ \href@noop {} {\emph {\bibinfo {booktitle} {Spectroscopic Ellipsometry}}}\ (\bibinfo  {publisher} {John Wiley \& Sons, Ltd},\ \bibinfo {year} {2007})\ Chap.\ \bibinfo {chapter} {2, 3, 4}, pp.\ \bibinfo {pages} {13--207}\BibitemShut {NoStop}%
\bibitem [{\citenamefont {Hauge}(1978)}]{Hauge:78}%
  \BibitemOpen
  \bibfield  {author} {\bibinfo {author} {\bibfnamefont {P.~S.}\ \bibnamefont {Hauge}},\ }\bibfield  {title} {\enquote {\bibinfo {title} {Mueller matrix ellipsometry with imperfect compensators},}\ }\href {https://doi.org/10.1364/JOSA.68.001519} {\bibfield  {journal} {\bibinfo  {journal} {J. Opt. Soc. Am.}\ }\textbf {\bibinfo {volume} {68}},\ \bibinfo {pages} {1519--1528} (\bibinfo {year} {1978})}\BibitemShut {NoStop}%
\bibitem [{\citenamefont {Stanislavchuk}\ \emph {et~al.}(2013)\citenamefont {Stanislavchuk}, \citenamefont {Kang}, \citenamefont {Rogers}, \citenamefont {Standard}, \citenamefont {Basistyy}, \citenamefont {Kotelyanskii}, \citenamefont {Nita}, \citenamefont {Zhou}, \citenamefont {Carr}, \citenamefont {Kotelyanskii},\ and\ \citenamefont {Sirenko}}]{10.1063/1.4789495}%
  \BibitemOpen
  \bibfield  {author} {\bibinfo {author} {\bibfnamefont {T.~N.}\ \bibnamefont {Stanislavchuk}}, \bibinfo {author} {\bibfnamefont {T.~D.}\ \bibnamefont {Kang}}, \bibinfo {author} {\bibfnamefont {P.~D.}\ \bibnamefont {Rogers}}, \bibinfo {author} {\bibfnamefont {E.~C.}\ \bibnamefont {Standard}}, \bibinfo {author} {\bibfnamefont {R.}~\bibnamefont {Basistyy}}, \bibinfo {author} {\bibfnamefont {A.~M.}\ \bibnamefont {Kotelyanskii}}, \bibinfo {author} {\bibfnamefont {G.}~\bibnamefont {Nita}}, \bibinfo {author} {\bibfnamefont {T.}~\bibnamefont {Zhou}}, \bibinfo {author} {\bibfnamefont {G.~L.}\ \bibnamefont {Carr}}, \bibinfo {author} {\bibfnamefont {M.}~\bibnamefont {Kotelyanskii}},\ and\ \bibinfo {author} {\bibfnamefont {A.~A.}\ \bibnamefont {Sirenko}},\ }\bibfield  {title} {\enquote {\bibinfo {title} {{Synchrotron radiation-based far-infrared spectroscopic ellipsometer with full Mueller-matrix capability}},}\ }\href@noop {} {\bibfield  {journal} {\bibinfo  {journal} {Review of Scientific Instruments}\ }\textbf
  {\bibinfo {volume} {84}},\ \bibinfo {pages} {023901} (\bibinfo {year} {2013})}\BibitemShut {NoStop}%
\bibitem [{\citenamefont {Pawar}\ \emph {et~al.}(2013)\citenamefont {Pawar}, \citenamefont {Sonawane}, \citenamefont {Erande},\ and\ \citenamefont {Derle}}]{PAWAR2013157}%
  \BibitemOpen
  \bibfield  {author} {\bibinfo {author} {\bibfnamefont {A.~Y.}\ \bibnamefont {Pawar}}, \bibinfo {author} {\bibfnamefont {D.~D.}\ \bibnamefont {Sonawane}}, \bibinfo {author} {\bibfnamefont {K.~B.}\ \bibnamefont {Erande}},\ and\ \bibinfo {author} {\bibfnamefont {D.~V.}\ \bibnamefont {Derle}},\ }\bibfield  {title} {\enquote {\bibinfo {title} {Terahertz technology and its applications},}\ }\href {https://doi.org/https://doi.org/10.1016/j.dit.2013.03.009} {\bibfield  {journal} {\bibinfo  {journal} {Drug Invention Today}\ }\textbf {\bibinfo {volume} {5}},\ \bibinfo {pages} {157--163} (\bibinfo {year} {2013})}\BibitemShut {NoStop}%
\bibitem [{\citenamefont {Ren}\ \emph {et~al.}(2009)\citenamefont {Ren}, \citenamefont {Pint}, \citenamefont {Booshehri}, \citenamefont {Rice}, \citenamefont {Wang}, \citenamefont {Hilton}, \citenamefont {Takeya}, \citenamefont {Kawayama}, \citenamefont {Tonouchi}, \citenamefont {Hauge},\ and\ \citenamefont {Kono}}]{doi:10.1021/nl900815s}%
  \BibitemOpen
  \bibfield  {author} {\bibinfo {author} {\bibfnamefont {L.}~\bibnamefont {Ren}}, \bibinfo {author} {\bibfnamefont {C.~L.}\ \bibnamefont {Pint}}, \bibinfo {author} {\bibfnamefont {L.~G.}\ \bibnamefont {Booshehri}}, \bibinfo {author} {\bibfnamefont {W.~D.}\ \bibnamefont {Rice}}, \bibinfo {author} {\bibfnamefont {X.}~\bibnamefont {Wang}}, \bibinfo {author} {\bibfnamefont {D.~J.}\ \bibnamefont {Hilton}}, \bibinfo {author} {\bibfnamefont {K.}~\bibnamefont {Takeya}}, \bibinfo {author} {\bibfnamefont {I.}~\bibnamefont {Kawayama}}, \bibinfo {author} {\bibfnamefont {M.}~\bibnamefont {Tonouchi}}, \bibinfo {author} {\bibfnamefont {R.~H.}\ \bibnamefont {Hauge}},\ and\ \bibinfo {author} {\bibfnamefont {J.}~\bibnamefont {Kono}},\ }\bibfield  {title} {\enquote {\bibinfo {title} {Carbon nanotube terahertz polarizer},}\ }\href {https://doi.org/10.1021/nl900815s} {\bibfield  {journal} {\bibinfo  {journal} {Nano Letters}\ }\textbf {\bibinfo {volume} {9}},\ \bibinfo {pages} {2610--2613} (\bibinfo {year} {2009})},\ \bibinfo
  {note} {pMID: 19492842},\ \Eprint {https://arxiv.org/abs/https://doi.org/10.1021/nl900815s} {https://doi.org/10.1021/nl900815s} \BibitemShut {NoStop}%
\bibitem [{\citenamefont {Ferraro}\ \emph {et~al.}(2016)\citenamefont {Ferraro}, \citenamefont {Zografopoulos}, \citenamefont {Missori}, \citenamefont {Peccianti}, \citenamefont {Caputo},\ and\ \citenamefont {Beccherelli}}]{Ferraro:16}%
  \BibitemOpen
  \bibfield  {author} {\bibinfo {author} {\bibfnamefont {A.}~\bibnamefont {Ferraro}}, \bibinfo {author} {\bibfnamefont {D.~C.}\ \bibnamefont {Zografopoulos}}, \bibinfo {author} {\bibfnamefont {M.}~\bibnamefont {Missori}}, \bibinfo {author} {\bibfnamefont {M.}~\bibnamefont {Peccianti}}, \bibinfo {author} {\bibfnamefont {R.}~\bibnamefont {Caputo}},\ and\ \bibinfo {author} {\bibfnamefont {R.}~\bibnamefont {Beccherelli}},\ }\bibfield  {title} {\enquote {\bibinfo {title} {Flexible terahertz wire grid polarizer with high extinction ratio and low loss},}\ }\href {https://doi.org/10.1364/OL.41.002009} {\bibfield  {journal} {\bibinfo  {journal} {Opt. Lett.}\ }\textbf {\bibinfo {volume} {41}},\ \bibinfo {pages} {2009--2012} (\bibinfo {year} {2016})}\BibitemShut {NoStop}%
\bibitem [{\citenamefont {Huang}\ \emph {et~al.}(2013)\citenamefont {Huang}, \citenamefont {Park}, \citenamefont {Parrott}, \citenamefont {Chan},\ and\ \citenamefont {Pickwell-MacPherson}}]{6357219}%
  \BibitemOpen
  \bibfield  {author} {\bibinfo {author} {\bibfnamefont {Z.}~\bibnamefont {Huang}}, \bibinfo {author} {\bibfnamefont {H.}~\bibnamefont {Park}}, \bibinfo {author} {\bibfnamefont {E.~P.~J.}\ \bibnamefont {Parrott}}, \bibinfo {author} {\bibfnamefont {H.~P.}\ \bibnamefont {Chan}},\ and\ \bibinfo {author} {\bibfnamefont {E.}~\bibnamefont {Pickwell-MacPherson}},\ }\bibfield  {title} {\enquote {\bibinfo {title} {Robust thin-film wire-grid thz polarizer fabricated via a low-cost approach},}\ }\href {https://doi.org/10.1109/LPT.2012.2228184} {\bibfield  {journal} {\bibinfo  {journal} {IEEE Photonics Technology Letters}\ }\textbf {\bibinfo {volume} {25}},\ \bibinfo {pages} {81--84} (\bibinfo {year} {2013})}\BibitemShut {NoStop}%
\bibitem [{\citenamefont {Lin}\ \emph {et~al.}(2011)\citenamefont {Lin}, \citenamefont {Wu}, \citenamefont {Hu}, \citenamefont {Zheng}, \citenamefont {Wu}, \citenamefont {Zhu}, \citenamefont {Xu}, \citenamefont {Jin},\ and\ \citenamefont {Lu}}]{10.1063/1.3626560}%
  \BibitemOpen
  \bibfield  {author} {\bibinfo {author} {\bibfnamefont {X.-w.}\ \bibnamefont {Lin}}, \bibinfo {author} {\bibfnamefont {J.-b.}\ \bibnamefont {Wu}}, \bibinfo {author} {\bibfnamefont {W.}~\bibnamefont {Hu}}, \bibinfo {author} {\bibfnamefont {Z.-g.}\ \bibnamefont {Zheng}}, \bibinfo {author} {\bibfnamefont {Z.-j.}\ \bibnamefont {Wu}}, \bibinfo {author} {\bibfnamefont {G.}~\bibnamefont {Zhu}}, \bibinfo {author} {\bibfnamefont {F.}~\bibnamefont {Xu}}, \bibinfo {author} {\bibfnamefont {B.-b.}\ \bibnamefont {Jin}},\ and\ \bibinfo {author} {\bibfnamefont {Y.-q.}\ \bibnamefont {Lu}},\ }\bibfield  {title} {\enquote {\bibinfo {title} {{Self-polarizing terahertz liquid crystal phase shifter}},}\ }\href {https://doi.org/10.1063/1.3626560} {\bibfield  {journal} {\bibinfo  {journal} {AIP Advances}\ }\textbf {\bibinfo {volume} {1}},\ \bibinfo {pages} {032133} (\bibinfo {year} {2011})},\ \Eprint {https://arxiv.org/abs/https://pubs.aip.org/aip/adv/article-pdf/doi/10.1063/1.3626560/12994921/032133\_1\_online.pdf}
  {https://pubs.aip.org/aip/adv/article-pdf/doi/10.1063/1.3626560/12994921/032133\_1\_online.pdf} \BibitemShut {NoStop}%
\bibitem [{\citenamefont {Reid}\ and\ \citenamefont {Fedosejevs}(2006)}]{Reid:06}%
  \BibitemOpen
  \bibfield  {author} {\bibinfo {author} {\bibfnamefont {M.}~\bibnamefont {Reid}}\ and\ \bibinfo {author} {\bibfnamefont {R.}~\bibnamefont {Fedosejevs}},\ }\bibfield  {title} {\enquote {\bibinfo {title} {Terahertz birefringence and attenuation properties of wood and paper},}\ }\href {https://doi.org/10.1364/AO.45.002766} {\bibfield  {journal} {\bibinfo  {journal} {Appl. Opt.}\ }\textbf {\bibinfo {volume} {45}},\ \bibinfo {pages} {2766--2772} (\bibinfo {year} {2006})}\BibitemShut {NoStop}%
\bibitem [{\citenamefont {Todoruk}, \citenamefont {Hartley},\ and\ \citenamefont {Reid}(2012)}]{6119245}%
  \BibitemOpen
  \bibfield  {author} {\bibinfo {author} {\bibfnamefont {T.~M.}\ \bibnamefont {Todoruk}}, \bibinfo {author} {\bibfnamefont {I.~D.}\ \bibnamefont {Hartley}},\ and\ \bibinfo {author} {\bibfnamefont {M.~E.}\ \bibnamefont {Reid}},\ }\bibfield  {title} {\enquote {\bibinfo {title} {Origin of birefringence in wood at terahertz frequencies},}\ }\href {https://doi.org/10.1109/TTHZ.2011.2177692} {\bibfield  {journal} {\bibinfo  {journal} {IEEE Transactions on Terahertz Science and Technology}\ }\textbf {\bibinfo {volume} {2}},\ \bibinfo {pages} {123--130} (\bibinfo {year} {2012})}\BibitemShut {NoStop}%
\bibitem [{\citenamefont {Hernandez-Serrano}, \citenamefont {Castro-Camus},\ and\ \citenamefont {Lopez-Mago}(2017)}]{Hernandez}%
  \BibitemOpen
  \bibfield  {author} {\bibinfo {author} {\bibfnamefont {A.~I.}\ \bibnamefont {Hernandez-Serrano}}, \bibinfo {author} {\bibfnamefont {E.}~\bibnamefont {Castro-Camus}},\ and\ \bibinfo {author} {\bibfnamefont {D.}~\bibnamefont {Lopez-Mago}},\ }\bibfield  {title} {\enquote {\bibinfo {title} {q-plate for the generation of terahertz cylindrical vector beams fabricated by 3d printing},}\ }\href {https://doi.org/10.1007/s10762-017-0396-8} {\bibfield  {journal} {\bibinfo  {journal} {Journal of Infrared, Millimeter, and Terahertz Waves}\ }\textbf {\bibinfo {volume} {38}},\ \bibinfo {pages} {938--944} (\bibinfo {year} {2017})}\BibitemShut {NoStop}%
\bibitem [{\citenamefont {Rohrbach}, \citenamefont {Kang},\ and\ \citenamefont {Feurer}(2021)}]{Rohrbach}%
  \BibitemOpen
  \bibfield  {author} {\bibinfo {author} {\bibfnamefont {D.}~\bibnamefont {Rohrbach}}, \bibinfo {author} {\bibfnamefont {B.~J.}\ \bibnamefont {Kang}},\ and\ \bibinfo {author} {\bibfnamefont {T.}~\bibnamefont {Feurer}},\ }\bibfield  {title} {\enquote {\bibinfo {title} {3d-printed thz wave- and phaseplates},}\ }\href {https://doi.org/10.1364/OE.433881} {\bibfield  {journal} {\bibinfo  {journal} {Opt. Express}\ }\textbf {\bibinfo {volume} {29}},\ \bibinfo {pages} {27160--27170} (\bibinfo {year} {2021})}\BibitemShut {NoStop}%
\bibitem [{\citenamefont {Castro-Camus}, \citenamefont {Koch},\ and\ \citenamefont {Hernandez-Serrano}(2020)}]{perspective}%
  \BibitemOpen
  \bibfield  {author} {\bibinfo {author} {\bibfnamefont {E.}~\bibnamefont {Castro-Camus}}, \bibinfo {author} {\bibfnamefont {M.}~\bibnamefont {Koch}},\ and\ \bibinfo {author} {\bibfnamefont {A.~I.}\ \bibnamefont {Hernandez-Serrano}},\ }\bibfield  {title} {\enquote {\bibinfo {title} {{Additive manufacture of photonic components for the terahertz band}},}\ }\href {https://doi.org/10.1063/1.5140270} {\bibfield  {journal} {\bibinfo  {journal} {Journal of Applied Physics}\ }\textbf {\bibinfo {volume} {127}},\ \bibinfo {pages} {210901} (\bibinfo {year} {2020})},\ \Eprint {https://arxiv.org/abs/https://pubs.aip.org/aip/jap/article-pdf/doi/10.1063/1.5140270/19997030/210901\_1\_1.5140270.pdf} {https://pubs.aip.org/aip/jap/article-pdf/doi/10.1063/1.5140270/19997030/210901\_1\_1.5140270.pdf} \BibitemShut {NoStop}%
\bibitem [{\citenamefont {Chen}\ and\ \citenamefont {Pickwell-MacPherson}(2022)}]{10.1063/5.0094056}%
  \BibitemOpen
  \bibfield  {author} {\bibinfo {author} {\bibfnamefont {X.}~\bibnamefont {Chen}}\ and\ \bibinfo {author} {\bibfnamefont {E.}~\bibnamefont {Pickwell-MacPherson}},\ }\bibfield  {title} {\enquote {\bibinfo {title} {{An introduction to terahertz time-domain spectroscopic ellipsometry}},}\ }\href {https://doi.org/10.1063/5.0094056} {\bibfield  {journal} {\bibinfo  {journal} {APL Photonics}\ }\textbf {\bibinfo {volume} {7}},\ \bibinfo {pages} {071101} (\bibinfo {year} {2022})},\ \Eprint {https://arxiv.org/abs/https://pubs.aip.org/aip/app/article-pdf/doi/10.1063/5.0094056/19854313/071101\_1\_5.0094056.pdf} {https://pubs.aip.org/aip/app/article-pdf/doi/10.1063/5.0094056/19854313/071101\_1\_5.0094056.pdf} \BibitemShut {NoStop}%
\bibitem [{\citenamefont {Guo}\ \emph {et~al.}(2019)\citenamefont {Guo}, \citenamefont {Zhang}, \citenamefont {Lyu}, \citenamefont {Zhang}, \citenamefont {Huang}, \citenamefont {Meng}, \citenamefont {Zhao},\ and\ \citenamefont {Yuan}}]{8735790}%
  \BibitemOpen
  \bibfield  {author} {\bibinfo {author} {\bibfnamefont {Q.}~\bibnamefont {Guo}}, \bibinfo {author} {\bibfnamefont {Y.}~\bibnamefont {Zhang}}, \bibinfo {author} {\bibfnamefont {Z.}~\bibnamefont {Lyu}}, \bibinfo {author} {\bibfnamefont {D.}~\bibnamefont {Zhang}}, \bibinfo {author} {\bibfnamefont {Y.}~\bibnamefont {Huang}}, \bibinfo {author} {\bibfnamefont {C.}~\bibnamefont {Meng}}, \bibinfo {author} {\bibfnamefont {Z.}~\bibnamefont {Zhao}},\ and\ \bibinfo {author} {\bibfnamefont {J.}~\bibnamefont {Yuan}},\ }\bibfield  {title} {\enquote {\bibinfo {title} {Thz time-domain spectroscopic ellipsometry with simultaneous measurements of orthogonal polarizations},}\ }\href {https://doi.org/10.1109/TTHZ.2019.2921200} {\bibfield  {journal} {\bibinfo  {journal} {IEEE Transactions on Terahertz Science and Technology}\ }\textbf {\bibinfo {volume} {9}},\ \bibinfo {pages} {422--429} (\bibinfo {year} {2019})}\BibitemShut {NoStop}%
\bibitem [{\citenamefont {Ketelsen}\ \emph {et~al.}(2022)\citenamefont {Ketelsen}, \citenamefont {Mästle}, \citenamefont {Liebermeister}, \citenamefont {Kohlhaas},\ and\ \citenamefont {Globisch}}]{app12083744}%
  \BibitemOpen
  \bibfield  {author} {\bibinfo {author} {\bibfnamefont {H.}~\bibnamefont {Ketelsen}}, \bibinfo {author} {\bibfnamefont {R.}~\bibnamefont {Mästle}}, \bibinfo {author} {\bibfnamefont {L.}~\bibnamefont {Liebermeister}}, \bibinfo {author} {\bibfnamefont {R.}~\bibnamefont {Kohlhaas}},\ and\ \bibinfo {author} {\bibfnamefont {B.}~\bibnamefont {Globisch}},\ }\bibfield  {title} {\enquote {\bibinfo {title} {Thz time-domain ellipsometer for material characterization and paint quality control with more than 5 thz bandwidth},}\ }\href {https://doi.org/10.3390/app12083744} {\bibfield  {journal} {\bibinfo  {journal} {Applied Sciences}\ }\textbf {\bibinfo {volume} {12}} (\bibinfo {year} {2022}),\ 10.3390/app12083744}\BibitemShut {NoStop}%
\bibitem [{\citenamefont {Mazaheri}, \citenamefont {Koral},\ and\ \citenamefont {Andreone}(2022)}]{Mazaheri2022}%
  \BibitemOpen
  \bibfield  {author} {\bibinfo {author} {\bibfnamefont {Z.}~\bibnamefont {Mazaheri}}, \bibinfo {author} {\bibfnamefont {C.}~\bibnamefont {Koral}},\ and\ \bibinfo {author} {\bibfnamefont {A.}~\bibnamefont {Andreone}},\ }\bibfield  {title} {\enquote {\bibinfo {title} {Accurate thz ellipsometry using calibration in time domain},}\ }\href {https://doi.org/10.1038/s41598-022-10804-w} {\bibfield  {journal} {\bibinfo  {journal} {Scientific Reports}\ }\textbf {\bibinfo {volume} {12}},\ \bibinfo {pages} {7342} (\bibinfo {year} {2022})}\BibitemShut {NoStop}%
\bibitem [{\citenamefont {Neshat}\ and\ \citenamefont {Armitage}(2012)}]{Neshat:12}%
  \BibitemOpen
  \bibfield  {author} {\bibinfo {author} {\bibfnamefont {M.}~\bibnamefont {Neshat}}\ and\ \bibinfo {author} {\bibfnamefont {N.~P.}\ \bibnamefont {Armitage}},\ }\bibfield  {title} {\enquote {\bibinfo {title} {Terahertz time-domain spectroscopic ellipsometry: instrumentation and calibration},}\ }\href {https://doi.org/10.1364/OE.20.029063} {\bibfield  {journal} {\bibinfo  {journal} {Opt. Express}\ }\textbf {\bibinfo {volume} {20}},\ \bibinfo {pages} {29063--29075} (\bibinfo {year} {2012})}\BibitemShut {NoStop}%
\bibitem [{\citenamefont {Matsumoto}\ \emph {et~al.}(2011)\citenamefont {Matsumoto}, \citenamefont {Hosokura}, \citenamefont {Nagashima},\ and\ \citenamefont {Hangyo}}]{Matsumoto:11}%
  \BibitemOpen
  \bibfield  {author} {\bibinfo {author} {\bibfnamefont {N.}~\bibnamefont {Matsumoto}}, \bibinfo {author} {\bibfnamefont {T.}~\bibnamefont {Hosokura}}, \bibinfo {author} {\bibfnamefont {T.}~\bibnamefont {Nagashima}},\ and\ \bibinfo {author} {\bibfnamefont {M.}~\bibnamefont {Hangyo}},\ }\bibfield  {title} {\enquote {\bibinfo {title} {Measurement of the dielectric constant of thin films by terahertz time-domain spectroscopic ellipsometry},}\ }\href {https://doi.org/10.1364/OL.36.000265} {\bibfield  {journal} {\bibinfo  {journal} {Opt. Lett.}\ }\textbf {\bibinfo {volume} {36}},\ \bibinfo {pages} {265--267} (\bibinfo {year} {2011})}\BibitemShut {NoStop}%
\bibitem [{\citenamefont {Xu}\ \emph {et~al.}(2020)\citenamefont {Xu}, \citenamefont {Bayati}, \citenamefont {Oguchi}, \citenamefont {Watanabe}, \citenamefont {Winebrenner},\ and\ \citenamefont {Arbab}}]{Xu:20}%
  \BibitemOpen
  \bibfield  {author} {\bibinfo {author} {\bibfnamefont {K.}~\bibnamefont {Xu}}, \bibinfo {author} {\bibfnamefont {E.}~\bibnamefont {Bayati}}, \bibinfo {author} {\bibfnamefont {K.}~\bibnamefont {Oguchi}}, \bibinfo {author} {\bibfnamefont {S.}~\bibnamefont {Watanabe}}, \bibinfo {author} {\bibfnamefont {D.~P.}\ \bibnamefont {Winebrenner}},\ and\ \bibinfo {author} {\bibfnamefont {M.~H.}\ \bibnamefont {Arbab}},\ }\bibfield  {title} {\enquote {\bibinfo {title} {Terahertz time-domain polarimetry (thz-tdp) based on the spinning e-o sampling technique: determination of precision and calibration},}\ }\href {https://doi.org/10.1364/OE.389651} {\bibfield  {journal} {\bibinfo  {journal} {Opt. Express}\ }\textbf {\bibinfo {volume} {28}},\ \bibinfo {pages} {13482--13496} (\bibinfo {year} {2020})}\BibitemShut {NoStop}%
\bibitem [{\citenamefont {Nagashima}\ and\ \citenamefont {Hangyo}(2001)}]{10.1063/1.1426258}%
  \BibitemOpen
  \bibfield  {author} {\bibinfo {author} {\bibfnamefont {T.}~\bibnamefont {Nagashima}}\ and\ \bibinfo {author} {\bibfnamefont {M.}~\bibnamefont {Hangyo}},\ }\bibfield  {title} {\enquote {\bibinfo {title} {{Measurement of complex optical constants of a highly doped Si wafer using terahertz ellipsometry}},}\ }\href {https://doi.org/10.1063/1.1426258} {\bibfield  {journal} {\bibinfo  {journal} {Applied Physics Letters}\ }\textbf {\bibinfo {volume} {79}},\ \bibinfo {pages} {3917--3919} (\bibinfo {year} {2001})},\ \Eprint {https://arxiv.org/abs/https://pubs.aip.org/aip/apl/article-pdf/79/24/3917/18562499/3917\_1\_online.pdf} {https://pubs.aip.org/aip/apl/article-pdf/79/24/3917/18562499/3917\_1\_online.pdf} \BibitemShut {NoStop}%
\bibitem [{\citenamefont {Mazaheri}\ \emph {et~al.}(2022)\citenamefont {Mazaheri}, \citenamefont {Koral}, \citenamefont {Andreone},\ and\ \citenamefont {Marino}}]{Mazaheri:22}%
  \BibitemOpen
  \bibfield  {author} {\bibinfo {author} {\bibfnamefont {Z.}~\bibnamefont {Mazaheri}}, \bibinfo {author} {\bibfnamefont {C.}~\bibnamefont {Koral}}, \bibinfo {author} {\bibfnamefont {A.}~\bibnamefont {Andreone}},\ and\ \bibinfo {author} {\bibfnamefont {A.}~\bibnamefont {Marino}},\ }\bibfield  {title} {\enquote {\bibinfo {title} {Terahertz time-domain ellipsometry: tutorial},}\ }\href {https://doi.org/10.1364/JOSAA.463969} {\bibfield  {journal} {\bibinfo  {journal} {J. Opt. Soc. Am. A}\ }\textbf {\bibinfo {volume} {39}},\ \bibinfo {pages} {1420--1433} (\bibinfo {year} {2022})}\BibitemShut {NoStop}%
\bibitem [{\citenamefont {Agulto}\ \emph {et~al.}(2022)\citenamefont {Agulto}, \citenamefont {Iwamoto}, \citenamefont {Mag-usara},\ and\ \citenamefont {Nakajima}}]{Agulto:22}%
  \BibitemOpen
  \bibfield  {author} {\bibinfo {author} {\bibfnamefont {V.~C.}\ \bibnamefont {Agulto}}, \bibinfo {author} {\bibfnamefont {T.}~\bibnamefont {Iwamoto}}, \bibinfo {author} {\bibfnamefont {V.~K.}\ \bibnamefont {Mag-usara}},\ and\ \bibinfo {author} {\bibfnamefont {M.}~\bibnamefont {Nakajima}},\ }\bibfield  {title} {\enquote {\bibinfo {title} {Development of terahertz time-domain rotating-analyzer ellipsometry},}\ }in\ \href {https://doi.org/10.1364/CLEOPR.2022.CTuP3C_02} {\emph {\bibinfo {booktitle} {Proceedings of the 2022 Conference on Lasers and Electro-Optics Pacific Rim}}}\ (\bibinfo  {publisher} {Optica Publishing Group},\ \bibinfo {year} {2022})\ p.\ \bibinfo {pages} {CTuP3C02}\BibitemShut {NoStop}%
\bibitem [{\citenamefont {Schubert}\ \emph {et~al.}(2022)\citenamefont {Schubert}, \citenamefont {Knight}, \citenamefont {Richter}, \citenamefont {Kühne}, \citenamefont {Stanishev}, \citenamefont {Ruder}, \citenamefont {Stokey}, \citenamefont {Korlacki}, \citenamefont {Irmscher}, \citenamefont {Neugebauer},\ and\ \citenamefont {Darakchieva}}]{10.1063/5.0082353}%
  \BibitemOpen
  \bibfield  {author} {\bibinfo {author} {\bibfnamefont {M.}~\bibnamefont {Schubert}}, \bibinfo {author} {\bibfnamefont {S.}~\bibnamefont {Knight}}, \bibinfo {author} {\bibfnamefont {S.}~\bibnamefont {Richter}}, \bibinfo {author} {\bibfnamefont {P.}~\bibnamefont {Kühne}}, \bibinfo {author} {\bibfnamefont {V.}~\bibnamefont {Stanishev}}, \bibinfo {author} {\bibfnamefont {A.}~\bibnamefont {Ruder}}, \bibinfo {author} {\bibfnamefont {M.}~\bibnamefont {Stokey}}, \bibinfo {author} {\bibfnamefont {R.}~\bibnamefont {Korlacki}}, \bibinfo {author} {\bibfnamefont {K.}~\bibnamefont {Irmscher}}, \bibinfo {author} {\bibfnamefont {P.}~\bibnamefont {Neugebauer}},\ and\ \bibinfo {author} {\bibfnamefont {V.}~\bibnamefont {Darakchieva}},\ }\bibfield  {title} {\enquote {\bibinfo {title} {Terahertz electron paramagnetic resonance generalized spectroscopic ellipsometry: The magnetic response of the nitrogen defect in 4h-sic},}\ }\href {https://doi.org/10.1063/5.0082353} {\bibfield  {journal} {\bibinfo  {journal} {Appl. Phys.
  Lett}\ }\textbf {\bibinfo {volume} {120}},\ \bibinfo {pages} {102101} (\bibinfo {year} {2022})},\ \Eprint {https://arxiv.org/abs/https://doi.org/10.1063/5.0082353} {https://doi.org/10.1063/5.0082353} \BibitemShut {NoStop}%
\bibitem [{\citenamefont {Kühne}\ \emph {et~al.}(2014)\citenamefont {Kühne}, \citenamefont {Herzinger}, \citenamefont {Schubert}, \citenamefont {Woollam},\ and\ \citenamefont {Hofmann}}]{10.1063/1.4889920}%
  \BibitemOpen
  \bibfield  {author} {\bibinfo {author} {\bibfnamefont {P.}~\bibnamefont {Kühne}}, \bibinfo {author} {\bibfnamefont {C.~M.}\ \bibnamefont {Herzinger}}, \bibinfo {author} {\bibfnamefont {M.}~\bibnamefont {Schubert}}, \bibinfo {author} {\bibfnamefont {J.~A.}\ \bibnamefont {Woollam}},\ and\ \bibinfo {author} {\bibfnamefont {T.}~\bibnamefont {Hofmann}},\ }\bibfield  {title} {\enquote {\bibinfo {title} {{Invited Article: An integrated mid-infrared, far-infrared, and terahertz optical Hall effect instrument}},}\ }\href {https://doi.org/10.1063/1.4889920} {\bibfield  {journal} {\bibinfo  {journal} {Review of Scientific Instruments}\ }\textbf {\bibinfo {volume} {85}},\ \bibinfo {pages} {071301} (\bibinfo {year} {2014})},\ \Eprint {https://arxiv.org/abs/https://pubs.aip.org/aip/rsi/article-pdf/doi/10.1063/1.4889920/14861969/071301\_1\_online.pdf} {https://pubs.aip.org/aip/rsi/article-pdf/doi/10.1063/1.4889920/14861969/071301\_1\_online.pdf} \BibitemShut {NoStop}%
\bibitem [{\citenamefont {Hofmann}\ \emph {et~al.}(2010)\citenamefont {Hofmann}, \citenamefont {Herzinger}, \citenamefont {Boosalis}, \citenamefont {Tiwald}, \citenamefont {Woollam},\ and\ \citenamefont {Schubert}}]{10.1063/1.3297902}%
  \BibitemOpen
  \bibfield  {author} {\bibinfo {author} {\bibfnamefont {T.}~\bibnamefont {Hofmann}}, \bibinfo {author} {\bibfnamefont {C.~M.}\ \bibnamefont {Herzinger}}, \bibinfo {author} {\bibfnamefont {A.}~\bibnamefont {Boosalis}}, \bibinfo {author} {\bibfnamefont {T.~E.}\ \bibnamefont {Tiwald}}, \bibinfo {author} {\bibfnamefont {J.~A.}\ \bibnamefont {Woollam}},\ and\ \bibinfo {author} {\bibfnamefont {M.}~\bibnamefont {Schubert}},\ }\bibfield  {title} {\enquote {\bibinfo {title} {{Variable-wavelength frequency-domain terahertz ellipsometry}},}\ }\href {https://doi.org/10.1063/1.3297902} {\bibfield  {journal} {\bibinfo  {journal} {Review of Scientific Instruments}\ }\textbf {\bibinfo {volume} {81}},\ \bibinfo {pages} {023101} (\bibinfo {year} {2010})},\ \Eprint {https://arxiv.org/abs/https://pubs.aip.org/aip/rsi/article-pdf/doi/10.1063/1.3297902/13929460/023101\_1\_online.pdf} {https://pubs.aip.org/aip/rsi/article-pdf/doi/10.1063/1.3297902/13929460/023101\_1\_online.pdf} \BibitemShut {NoStop}%
\bibitem [{\citenamefont {Klenner}\ \emph {et~al.}(2016)\citenamefont {Klenner}, \citenamefont {Zech}, \citenamefont {Hülsmann}, \citenamefont {Kühn}, \citenamefont {Schlechtweg},\ and\ \citenamefont {Ambacher}}]{7534823}%
  \BibitemOpen
  \bibfield  {author} {\bibinfo {author} {\bibfnamefont {M.}~\bibnamefont {Klenner}}, \bibinfo {author} {\bibfnamefont {C.}~\bibnamefont {Zech}}, \bibinfo {author} {\bibfnamefont {A.}~\bibnamefont {Hülsmann}}, \bibinfo {author} {\bibfnamefont {J.}~\bibnamefont {Kühn}}, \bibinfo {author} {\bibfnamefont {M.}~\bibnamefont {Schlechtweg}},\ and\ \bibinfo {author} {\bibfnamefont {O.}~\bibnamefont {Ambacher}},\ }\bibfield  {title} {\enquote {\bibinfo {title} {Spectroscopic measurement of material properties using an improved millimeter-wave ellipsometer based on metallic substrates},}\ }\href {https://doi.org/10.1109/TIM.2016.2594022} {\bibfield  {journal} {\bibinfo  {journal} {IEEE Transactions on Instrumentation and Measurement}\ }\textbf {\bibinfo {volume} {65}},\ \bibinfo {pages} {2551--2559} (\bibinfo {year} {2016})}\BibitemShut {NoStop}%
\bibitem [{\citenamefont {Kühne}\ \emph {et~al.}(2018)\citenamefont {Kühne}, \citenamefont {Armakavicius}, \citenamefont {Stanishev}, \citenamefont {Herzinger}, \citenamefont {Schubert},\ and\ \citenamefont {Darakchieva}}]{2018LuEllipsometer}%
  \BibitemOpen
  \bibfield  {author} {\bibinfo {author} {\bibfnamefont {P.}~\bibnamefont {Kühne}}, \bibinfo {author} {\bibfnamefont {N.}~\bibnamefont {Armakavicius}}, \bibinfo {author} {\bibfnamefont {V.}~\bibnamefont {Stanishev}}, \bibinfo {author} {\bibfnamefont {C.~M.}\ \bibnamefont {Herzinger}}, \bibinfo {author} {\bibfnamefont {M.}~\bibnamefont {Schubert}},\ and\ \bibinfo {author} {\bibfnamefont {V.}~\bibnamefont {Darakchieva}},\ }\bibfield  {title} {\enquote {\bibinfo {title} {Advanced terahertz frequency-domain ellipsometry instrumentation for in situ and ex situ applications},}\ }\href {https://doi.org/10.1109/TTHZ.2018.2814347} {\bibfield  {journal} {\bibinfo  {journal} {IEEE Transactions on Terahertz Science and Technology}\ }\textbf {\bibinfo {volume} {8}},\ \bibinfo {pages} {257--270} (\bibinfo {year} {2018})}\BibitemShut {NoStop}%
\bibitem [{\citenamefont {Hofmann}\ \emph {et~al.}(2011)\citenamefont {Hofmann}, \citenamefont {Herzinger}, \citenamefont {Tedesco}, \citenamefont {Gaskill}, \citenamefont {Woollam},\ and\ \citenamefont {Schubert}}]{HOFMANN20112593}%
  \BibitemOpen
  \bibfield  {author} {\bibinfo {author} {\bibfnamefont {T.}~\bibnamefont {Hofmann}}, \bibinfo {author} {\bibfnamefont {C.}~\bibnamefont {Herzinger}}, \bibinfo {author} {\bibfnamefont {J.}~\bibnamefont {Tedesco}}, \bibinfo {author} {\bibfnamefont {D.}~\bibnamefont {Gaskill}}, \bibinfo {author} {\bibfnamefont {J.}~\bibnamefont {Woollam}},\ and\ \bibinfo {author} {\bibfnamefont {M.}~\bibnamefont {Schubert}},\ }\bibfield  {title} {\enquote {\bibinfo {title} {Terahertz ellipsometry and terahertz optical-hall effect},}\ }\href {https://doi.org/https://doi.org/10.1016/j.tsf.2010.11.069} {\bibfield  {journal} {\bibinfo  {journal} {Thin Solid Films}\ }\textbf {\bibinfo {volume} {519}},\ \bibinfo {pages} {2593--2600} (\bibinfo {year} {2011})},\ \bibinfo {note} {5th International Conference on Spectroscopic Ellipsometry (ICSE-V)}\BibitemShut {NoStop}%
\bibitem [{\citenamefont {Belyaeva}\ \emph {et~al.}(2021)\citenamefont {Belyaeva}, \citenamefont {Galuza}, \citenamefont {Kolenov},\ and\ \citenamefont {Mizrakhy}}]{Belyaeva2021-in}%
  \BibitemOpen
  \bibfield  {author} {\bibinfo {author} {\bibfnamefont {A.}~\bibnamefont {Belyaeva}}, \bibinfo {author} {\bibfnamefont {A.}~\bibnamefont {Galuza}}, \bibinfo {author} {\bibfnamefont {I.}~\bibnamefont {Kolenov}},\ and\ \bibinfo {author} {\bibfnamefont {S.}~\bibnamefont {Mizrakhy}},\ }\bibfield  {title} {\enquote {\bibinfo {title} {Developments in terahertz ellipsometry: Portable spectroscopic {Quasi-Optical} {Ellipsometer-Reflectometer} and its applications},}\ }\href@noop {} {\bibfield  {journal} {\bibinfo  {journal} {Journal of Infrared, Millimeter, and Terahertz Waves}\ }\textbf {\bibinfo {volume} {42}},\ \bibinfo {pages} {130--153} (\bibinfo {year} {2021})}\BibitemShut {NoStop}%
\bibitem [{\citenamefont {Hesler}\ \emph {et~al.}(2008)\citenamefont {Hesler}, \citenamefont {Liu}, \citenamefont {Xu}, \citenamefont {Duan},\ and\ \citenamefont {Weikle}}]{QOD1}%
  \BibitemOpen
  \bibfield  {author} {\bibinfo {author} {\bibfnamefont {J.~L.}\ \bibnamefont {Hesler}}, \bibinfo {author} {\bibfnamefont {L.}~\bibnamefont {Liu}}, \bibinfo {author} {\bibfnamefont {H.}~\bibnamefont {Xu}}, \bibinfo {author} {\bibfnamefont {Y.}~\bibnamefont {Duan}},\ and\ \bibinfo {author} {\bibfnamefont {R.~M.}\ \bibnamefont {Weikle}},\ }\href {https://doi.org/10.1109/ICIMW.2008.4665521} {\enquote {\bibinfo {title} {The development of quasi-optical thz detectors},}\ } (\bibinfo {year} {2008})\BibitemShut {NoStop}%
\bibitem [{\citenamefont {Liu}\ \emph {et~al.}(2010)\citenamefont {Liu}, \citenamefont {Hesler}, \citenamefont {Xu}, \citenamefont {Lichtenberger},\ and\ \citenamefont {Weikle}}]{QOD2}%
  \BibitemOpen
  \bibfield  {author} {\bibinfo {author} {\bibfnamefont {L.}~\bibnamefont {Liu}}, \bibinfo {author} {\bibfnamefont {J.~L.}\ \bibnamefont {Hesler}}, \bibinfo {author} {\bibfnamefont {H.}~\bibnamefont {Xu}}, \bibinfo {author} {\bibfnamefont {A.~W.}\ \bibnamefont {Lichtenberger}},\ and\ \bibinfo {author} {\bibfnamefont {R.~M.}\ \bibnamefont {Weikle}},\ }\bibfield  {title} {\enquote {\bibinfo {title} {A broadband quasi-optical terahertz detector utilizing a zero bias schottky diode},}\ }\href {https://doi.org/10.1109/LMWC.2010.2055553} {\bibfield  {journal} {\bibinfo  {journal} {IEEE Microwave and Wireless Components Letters}\ }\textbf {\bibinfo {volume} {20}},\ \bibinfo {pages} {504--506} (\bibinfo {year} {2010})}\BibitemShut {NoStop}%
\bibitem [{\citenamefont {Eaton}, \citenamefont {Eaton},\ and\ \citenamefont {Salikhov}(1998)}]{doi:10.1142/3624}%
  \BibitemOpen
  \bibfield  {author} {\bibinfo {author} {\bibfnamefont {G.~R.}\ \bibnamefont {Eaton}}, \bibinfo {author} {\bibfnamefont {S.~S.}\ \bibnamefont {Eaton}},\ and\ \bibinfo {author} {\bibfnamefont {K.~M.}\ \bibnamefont {Salikhov}},\ }\href {https://doi.org/10.1142/3624} {\emph {\bibinfo {title} {Foundations of Modern EPR}}}\ (\bibinfo  {publisher} {WORLD SCIENTIFIC},\ \bibinfo {year} {1998})\ \Eprint {https://arxiv.org/abs/https://www.worldscientific.com/doi/pdf/10.1142/3624} {https://www.worldscientific.com/doi/pdf/10.1142/3624} \BibitemShut {NoStop}%
\bibitem [{\citenamefont {Weil}\ and\ \citenamefont {Bolton}(2007)}]{WeilBolton2007}%
  \BibitemOpen
  \bibfield  {author} {\bibinfo {author} {\bibfnamefont {J.~A.}\ \bibnamefont {Weil}}\ and\ \bibinfo {author} {\bibfnamefont {J.~R.}\ \bibnamefont {Bolton}},\ }\href@noop {} {\emph {\bibinfo {title} {ELECTRON PARAMAGNETIC RESONANCE Elementary Theory and Practical Applications}}}\ (\bibinfo  {publisher} {John Wiley \& Sons},\ \bibinfo {address} {Hobokken NJ},\ \bibinfo {year} {2007})\BibitemShut {NoStop}%
\bibitem [{\citenamefont {Al'tshuler}\ and\ \citenamefont {Kozyrev}(1964)}]{al1964electron}%
  \BibitemOpen
  \bibfield  {author} {\bibinfo {author} {\bibfnamefont {S.}~\bibnamefont {Al'tshuler}}\ and\ \bibinfo {author} {\bibfnamefont {B.}~\bibnamefont {Kozyrev}},\ }\bibfield  {title} {\enquote {\bibinfo {title} {Electron paramagnetic resonance},}\ }\href@noop {} {\  (\bibinfo {year} {1964})}\BibitemShut {NoStop}%
\bibitem [{\citenamefont {Höfer}(2009)}]{höfer}%
  \BibitemOpen
  \bibfield  {author} {\bibinfo {author} {\bibfnamefont {P.}~\bibnamefont {Höfer}},\ }\enquote {\bibinfo {title} {Basic experimental methods in continuous wave electron paramagnetic resonance},}\ in\ \href {https://doi.org/https://doi.org/10.1002/9780470432235.ch2} {\emph {\bibinfo {booktitle} {Electron Paramagnetic Resonance}}}\ (\bibinfo  {publisher} {John Wiley $\&$ Sons, Ltd},\ \bibinfo {year} {2009})\ Chap.~\bibinfo {chapter} {2}, pp.\ \bibinfo {pages} {37--82},\ \Eprint {https://arxiv.org/abs/https://onlinelibrary.wiley.com/doi/pdf/10.1002/9780470432235.ch2} {https://onlinelibrary.wiley.com/doi/pdf/10.1002/9780470432235.ch2} \BibitemShut {NoStop}%
\bibitem [{\citenamefont {JOYCE}\ and\ \citenamefont {RICHARDS}(1969)}]{Richards1}%
  \BibitemOpen
  \bibfield  {author} {\bibinfo {author} {\bibfnamefont {R.~R.}\ \bibnamefont {JOYCE}}\ and\ \bibinfo {author} {\bibfnamefont {P.~L.}\ \bibnamefont {RICHARDS}},\ }\bibfield  {title} {\enquote {\bibinfo {title} {Far-infrared spectra of {${\mathrm{Al}}_{2}$${\mathrm{O}}_{3}$} doped with {Ti}, {V}, and {Cr}},}\ }\href {https://doi.org/10.1103/PhysRev.179.375} {\bibfield  {journal} {\bibinfo  {journal} {Phys. Rev.}\ }\textbf {\bibinfo {volume} {179}},\ \bibinfo {pages} {375--380} (\bibinfo {year} {1969})}\BibitemShut {NoStop}%
\bibitem [{\citenamefont {Brackett}, \citenamefont {Richards},\ and\ \citenamefont {Caughey}(1971)}]{Richards2}%
  \BibitemOpen
  \bibfield  {author} {\bibinfo {author} {\bibfnamefont {G.~C.}\ \bibnamefont {Brackett}}, \bibinfo {author} {\bibfnamefont {P.~L.}\ \bibnamefont {Richards}},\ and\ \bibinfo {author} {\bibfnamefont {W.~S.}\ \bibnamefont {Caughey}},\ }\bibfield  {title} {\enquote {\bibinfo {title} {{Far‐Infrared Magnetic Resonance in Fe(III) and Mn(III) Porphyrins, Myoglobin, Hemoglobin, Ferrichrome A, and Fe(III) Dithiocarbamates}},}\ }\href {https://doi.org/10.1063/1.1674688} {\bibfield  {journal} {\bibinfo  {journal} {The Journal of Chemical Physics}\ }\textbf {\bibinfo {volume} {54}},\ \bibinfo {pages} {4383--4401} (\bibinfo {year} {1971})},\ \Eprint {https://arxiv.org/abs/https://pubs.aip.org/aip/jcp/article-pdf/54/10/4383/18873919/4383\_1\_online.pdf} {https://pubs.aip.org/aip/jcp/article-pdf/54/10/4383/18873919/4383\_1\_online.pdf} \BibitemShut {NoStop}%
\bibitem [{\citenamefont {Champion}\ and\ \citenamefont {Sievers}(1980)}]{Sievers}%
  \BibitemOpen
  \bibfield  {author} {\bibinfo {author} {\bibfnamefont {P.~M.}\ \bibnamefont {Champion}}\ and\ \bibinfo {author} {\bibfnamefont {A.~J.}\ \bibnamefont {Sievers}},\ }\bibfield  {title} {\enquote {\bibinfo {title} {{Far infrared magnetic resonance of deoxyhemoglobin and deoxymyoglobin}},}\ }\href {https://doi.org/10.1063/1.439356} {\bibfield  {journal} {\bibinfo  {journal} {The Journal of Chemical Physics}\ }\textbf {\bibinfo {volume} {72}},\ \bibinfo {pages} {1569--1582} (\bibinfo {year} {1980})},\ \Eprint {https://arxiv.org/abs/https://pubs.aip.org/aip/jcp/article-pdf/72/3/1569/18921379/1569\_1\_online.pdf} {https://pubs.aip.org/aip/jcp/article-pdf/72/3/1569/18921379/1569\_1\_online.pdf} \BibitemShut {NoStop}%
\bibitem [{\citenamefont {van Slageren}\ \emph {et~al.}(2003)\citenamefont {van Slageren}, \citenamefont {Vongtragool}, \citenamefont {Gorshunov}, \citenamefont {Mukhin}, \citenamefont {Karl}, \citenamefont {Krzystek}, \citenamefont {Telser}, \citenamefont {Müller}, \citenamefont {Sangregorio}, \citenamefont {Gatteschi},\ and\ \citenamefont {Dressel}}]{B305328H}%
  \BibitemOpen
  \bibfield  {author} {\bibinfo {author} {\bibfnamefont {J.}~\bibnamefont {van Slageren}}, \bibinfo {author} {\bibfnamefont {S.}~\bibnamefont {Vongtragool}}, \bibinfo {author} {\bibfnamefont {B.}~\bibnamefont {Gorshunov}}, \bibinfo {author} {\bibfnamefont {A.~A.}\ \bibnamefont {Mukhin}}, \bibinfo {author} {\bibfnamefont {N.}~\bibnamefont {Karl}}, \bibinfo {author} {\bibfnamefont {J.}~\bibnamefont {Krzystek}}, \bibinfo {author} {\bibfnamefont {J.}~\bibnamefont {Telser}}, \bibinfo {author} {\bibfnamefont {A.}~\bibnamefont {Müller}}, \bibinfo {author} {\bibfnamefont {C.}~\bibnamefont {Sangregorio}}, \bibinfo {author} {\bibfnamefont {D.}~\bibnamefont {Gatteschi}},\ and\ \bibinfo {author} {\bibfnamefont {M.}~\bibnamefont {Dressel}},\ }\bibfield  {title} {\enquote {\bibinfo {title} {Frequency-domain magnetic resonance spectroscopy of molecular magnetic materials},}\ }\href {https://doi.org/10.1039/B305328H} {\bibfield  {journal} {\bibinfo  {journal} {Phys. Chem. Chem. Phys.}\ }\textbf {\bibinfo {volume} {5}},\
  \bibinfo {pages} {3837--3843} (\bibinfo {year} {2003})}\BibitemShut {NoStop}%
\bibitem [{\citenamefont {Sojka}\ \emph {et~al.}(2020)\citenamefont {Sojka}, \citenamefont {Šedivý}, \citenamefont {Laguta}, \citenamefont {Marko}, \citenamefont {Santana},\ and\ \citenamefont {Neugebauer}}]{HFEPR}%
  \BibitemOpen
  \bibfield  {author} {\bibinfo {author} {\bibfnamefont {A.}~\bibnamefont {Sojka}}, \bibinfo {author} {\bibfnamefont {M.}~\bibnamefont {Šedivý}}, \bibinfo {author} {\bibfnamefont {O.}~\bibnamefont {Laguta}}, \bibinfo {author} {\bibfnamefont {A.}~\bibnamefont {Marko}}, \bibinfo {author} {\bibfnamefont {V.~T.}\ \bibnamefont {Santana}},\ and\ \bibinfo {author} {\bibfnamefont {P.}~\bibnamefont {Neugebauer}},\ }\bibfield  {title} {\enquote {\bibinfo {title} {{High-frequency EPR: current state and perspectives}},}\ }in\ \href {https://doi.org/10.1039/9781839162534-00214} {\emph {\bibinfo {booktitle} {{Electron Paramagnetic Resonance: Volume 27}}}}\ (\bibinfo  {publisher} {The Royal Society of Chemistry},\ \bibinfo {year} {2020})\BibitemShut {NoStop}%
\bibitem [{\citenamefont {Nehrkorn}\ \emph {et~al.}(2017)\citenamefont {Nehrkorn}, \citenamefont {Holldack}, \citenamefont {Bittl},\ and\ \citenamefont {Schnegg}}]{Schnegg}%
  \BibitemOpen
  \bibfield  {author} {\bibinfo {author} {\bibfnamefont {J.}~\bibnamefont {Nehrkorn}}, \bibinfo {author} {\bibfnamefont {K.}~\bibnamefont {Holldack}}, \bibinfo {author} {\bibfnamefont {R.}~\bibnamefont {Bittl}},\ and\ \bibinfo {author} {\bibfnamefont {A.}~\bibnamefont {Schnegg}},\ }\bibfield  {title} {\enquote {\bibinfo {title} {Recent progress in synchrotron-based frequency-domain fourier-transform thz-epr},}\ }\href {https://doi.org/10.1016/j.jmr.2017.04.001} {\bibfield  {journal} {\bibinfo  {journal} {Journal of Magnetic Resonance}\ }\textbf {\bibinfo {volume} {280}},\ \bibinfo {pages} {10--19} (\bibinfo {year} {2017})},\ \bibinfo {note} {copyright © 2017 Elsevier Inc. All rights reserved.}\BibitemShut {Stop}%
\bibitem [{\citenamefont {Neugebauer}\ \emph {et~al.}(2018)\citenamefont {Neugebauer}, \citenamefont {Bloos}, \citenamefont {Marx}, \citenamefont {Lutz}, \citenamefont {Kern}, \citenamefont {Aguilà}, \citenamefont {Vaverka}, \citenamefont {Laguta}, \citenamefont {Dietrich}, \citenamefont {Clérac},\ and\ \citenamefont {van Slageren}}]{Neugebauer}%
  \BibitemOpen
  \bibfield  {author} {\bibinfo {author} {\bibfnamefont {P.}~\bibnamefont {Neugebauer}}, \bibinfo {author} {\bibfnamefont {D.}~\bibnamefont {Bloos}}, \bibinfo {author} {\bibfnamefont {R.}~\bibnamefont {Marx}}, \bibinfo {author} {\bibfnamefont {P.}~\bibnamefont {Lutz}}, \bibinfo {author} {\bibfnamefont {M.}~\bibnamefont {Kern}}, \bibinfo {author} {\bibfnamefont {D.}~\bibnamefont {Aguilà}}, \bibinfo {author} {\bibfnamefont {J.}~\bibnamefont {Vaverka}}, \bibinfo {author} {\bibfnamefont {O.}~\bibnamefont {Laguta}}, \bibinfo {author} {\bibfnamefont {C.}~\bibnamefont {Dietrich}}, \bibinfo {author} {\bibfnamefont {R.}~\bibnamefont {Clérac}},\ and\ \bibinfo {author} {\bibfnamefont {J.}~\bibnamefont {van Slageren}},\ }\bibfield  {title} {\enquote {\bibinfo {title} {Ultra-broadband epr spectroscopy in field and frequency domains},}\ }\href {https://doi.org/10.1039/C7CP07443C} {\bibfield  {journal} {\bibinfo  {journal} {Phys. Chem. Chem. Phys.}\ }\textbf {\bibinfo {volume} {20}},\ \bibinfo {pages} {15528--15534}
  (\bibinfo {year} {2018})}\BibitemShut {NoStop}%
\bibitem [{\citenamefont {Khatri}\ \emph {et~al.}(2023)\citenamefont {Khatri}, \citenamefont {Fritjofson}, \citenamefont {Hanson-Flores}, \citenamefont {Kwon},\ and\ \citenamefont {Del~Barco}}]{Barco}%
  \BibitemOpen
  \bibfield  {author} {\bibinfo {author} {\bibfnamefont {G.}~\bibnamefont {Khatri}}, \bibinfo {author} {\bibfnamefont {G.}~\bibnamefont {Fritjofson}}, \bibinfo {author} {\bibfnamefont {J.}~\bibnamefont {Hanson-Flores}}, \bibinfo {author} {\bibfnamefont {J.}~\bibnamefont {Kwon}},\ and\ \bibinfo {author} {\bibfnamefont {E.}~\bibnamefont {Del~Barco}},\ }\bibfield  {title} {\enquote {\bibinfo {title} {{A 220 GHz–1.1 THz continuous frequency and polarization tunable quasi-optical electron paramagnetic resonance spectroscopic system}},}\ }\href {https://doi.org/10.1063/5.0107237} {\bibfield  {journal} {\bibinfo  {journal} {Review of Scientific Instruments}\ }\textbf {\bibinfo {volume} {94}},\ \bibinfo {pages} {034714} (\bibinfo {year} {2023})},\ \Eprint {https://arxiv.org/abs/https://pubs.aip.org/aip/rsi/article-pdf/doi/10.1063/5.0107237/16797752/034714\_1\_online.pdf} {https://pubs.aip.org/aip/rsi/article-pdf/doi/10.1063/5.0107237/16797752/034714\_1\_online.pdf} \BibitemShut {NoStop}%
\bibitem [{\citenamefont {Cho}, \citenamefont {Stepanov},\ and\ \citenamefont {Takahashi}(2014)}]{tdsepr1}%
  \BibitemOpen
  \bibfield  {author} {\bibinfo {author} {\bibfnamefont {F.~H.}\ \bibnamefont {Cho}}, \bibinfo {author} {\bibfnamefont {V.}~\bibnamefont {Stepanov}},\ and\ \bibinfo {author} {\bibfnamefont {S.}~\bibnamefont {Takahashi}},\ }\bibfield  {title} {\enquote {\bibinfo {title} {{A high-frequency electron paramagnetic resonance spectrometer for multi-dimensional, multi-frequency, and multi-phase pulsed measurements}},}\ }\href {https://doi.org/10.1063/1.4889873} {\bibfield  {journal} {\bibinfo  {journal} {Review of Scientific Instruments}\ }\textbf {\bibinfo {volume} {85}},\ \bibinfo {pages} {075110} (\bibinfo {year} {2014})},\ \Eprint {https://arxiv.org/abs/https://pubs.aip.org/aip/rsi/article-pdf/doi/10.1063/1.4889873/14865052/075110\_1\_online.pdf} {https://pubs.aip.org/aip/rsi/article-pdf/doi/10.1063/1.4889873/14865052/075110\_1\_online.pdf} \BibitemShut {NoStop}%
\bibitem [{\citenamefont {Lu}\ \emph {et~al.}(2017)\citenamefont {Lu}, \citenamefont {Ozel}, \citenamefont {Belvin}, \citenamefont {Li}, \citenamefont {Skorupskii}, \citenamefont {Sun}, \citenamefont {Ofori-Okai}, \citenamefont {Dincă}, \citenamefont {Gedik},\ and\ \citenamefont {Nelson}}]{tdsepr2}%
  \BibitemOpen
  \bibfield  {author} {\bibinfo {author} {\bibfnamefont {J.}~\bibnamefont {Lu}}, \bibinfo {author} {\bibfnamefont {I.~O.}\ \bibnamefont {Ozel}}, \bibinfo {author} {\bibfnamefont {C.~A.}\ \bibnamefont {Belvin}}, \bibinfo {author} {\bibfnamefont {X.}~\bibnamefont {Li}}, \bibinfo {author} {\bibfnamefont {G.}~\bibnamefont {Skorupskii}}, \bibinfo {author} {\bibfnamefont {L.}~\bibnamefont {Sun}}, \bibinfo {author} {\bibfnamefont {B.~K.}\ \bibnamefont {Ofori-Okai}}, \bibinfo {author} {\bibfnamefont {M.}~\bibnamefont {Dincă}}, \bibinfo {author} {\bibfnamefont {N.}~\bibnamefont {Gedik}},\ and\ \bibinfo {author} {\bibfnamefont {K.~A.}\ \bibnamefont {Nelson}},\ }\bibfield  {title} {\enquote {\bibinfo {title} {Rapid and precise determination of zero-field splittings by terahertz time-domain electron paramagnetic resonance spectroscopy},}\ }\href {https://doi.org/10.1039/C7SC00830A} {\bibfield  {journal} {\bibinfo  {journal} {Chem. Sci.}\ }\textbf {\bibinfo {volume} {8}},\ \bibinfo {pages} {7312--7323} (\bibinfo {year}
  {2017})}\BibitemShut {NoStop}%
\bibitem [{\citenamefont {Jackson}(1999)}]{JacksonCED}%
  \BibitemOpen
  \bibfield  {author} {\bibinfo {author} {\bibfnamefont {J.~D.}\ \bibnamefont {Jackson}},\ }\href {https://doi.org/10.1142/3624} {\emph {\bibinfo {title} {Classical Electrodynamics}}},\ \bibinfo {edition} {3rd}\ ed.\ (\bibinfo  {publisher} {John Wiley \& Sons},\ \bibinfo {year} {1999})\BibitemShut {NoStop}%
\bibitem [{\citenamefont {K\"{u}hne}\ \emph {et~al.}(2014)\citenamefont {K\"{u}hne}, \citenamefont {Herzinger}, \citenamefont {Schubert}, \citenamefont {Woollam},\ and\ \citenamefont {Hofmann}}]{doi:10.1063/1.4889920}%
  \BibitemOpen
  \bibfield  {author} {\bibinfo {author} {\bibfnamefont {P.}~\bibnamefont {K\"{u}hne}}, \bibinfo {author} {\bibfnamefont {C.~M.}\ \bibnamefont {Herzinger}}, \bibinfo {author} {\bibfnamefont {M.}~\bibnamefont {Schubert}}, \bibinfo {author} {\bibfnamefont {J.~A.}\ \bibnamefont {Woollam}},\ and\ \bibinfo {author} {\bibfnamefont {T.}~\bibnamefont {Hofmann}},\ }\bibfield  {title} {\enquote {\bibinfo {title} {Invited article: An integrated mid-infrared, far-infrared, and terahertz optical hall effect instrument},}\ }\href@noop {} {\bibfield  {journal} {\bibinfo  {journal} {Rev. Sci. Instr.}\ }\textbf {\bibinfo {volume} {85}},\ \bibinfo {pages} {071301} (\bibinfo {year} {2014})}\BibitemShut {NoStop}%
\bibitem [{\citenamefont {Schubert}(2004)}]{Schubert04}%
  \BibitemOpen
  \bibfield  {author} {\bibinfo {author} {\bibfnamefont {M.}~\bibnamefont {Schubert}},\ }\bibfield  {title} {\enquote {\bibinfo {title} {Theory and application of generalized ellipsometry},}\ }in\ \href@noop {} {\emph {\bibinfo {booktitle} {Handbook of Ellipsometry}}},\ \bibinfo {editor} {edited by\ \bibinfo {editor} {\bibfnamefont {E.}~\bibnamefont {Irene}}\ and\ \bibinfo {editor} {\bibfnamefont {H.}~\bibnamefont {Tompkins}}}\ (\bibinfo  {publisher} {William Andrew Publishing},\ \bibinfo {year} {2004})\BibitemShut {NoStop}%
\bibitem [{\citenamefont {Schubert}\ \emph {et~al.}(2016)\citenamefont {Schubert}, \citenamefont {K\"{u}hne}, \citenamefont {Darakchieva},\ and\ \citenamefont {Hofmann}}]{Schubert:16}%
  \BibitemOpen
  \bibfield  {author} {\bibinfo {author} {\bibfnamefont {M.}~\bibnamefont {Schubert}}, \bibinfo {author} {\bibfnamefont {P.}~\bibnamefont {K\"{u}hne}}, \bibinfo {author} {\bibfnamefont {V.}~\bibnamefont {Darakchieva}},\ and\ \bibinfo {author} {\bibfnamefont {T.}~\bibnamefont {Hofmann}},\ }\bibfield  {title} {\enquote {\bibinfo {title} {Optical hall effect---model description: tutorial},}\ }\href {https://doi.org/10.1364/JOSAA.33.001553} {\bibfield  {journal} {\bibinfo  {journal} {J. Opt. Soc. Am. A}\ }\textbf {\bibinfo {volume} {33}},\ \bibinfo {pages} {1553--1568} (\bibinfo {year} {2016})}\BibitemShut {NoStop}%
\bibitem [{\citenamefont {Schubert}, \citenamefont {Hofmann},\ and\ \citenamefont {Herzinger}(2003)}]{Schubert:03}%
  \BibitemOpen
  \bibfield  {author} {\bibinfo {author} {\bibfnamefont {M.}~\bibnamefont {Schubert}}, \bibinfo {author} {\bibfnamefont {T.}~\bibnamefont {Hofmann}},\ and\ \bibinfo {author} {\bibfnamefont {C.~M.}\ \bibnamefont {Herzinger}},\ }\bibfield  {title} {\enquote {\bibinfo {title} {Generalized far-infrared magneto-optic ellipsometry for semiconductor layer structures: determination of free-carrier effective-mass, mobility, and concentration parameters in n-type {GaAs}},}\ }\href {https://doi.org/10.1364/JOSAA.20.000347} {\bibfield  {journal} {\bibinfo  {journal} {J. Opt. Soc. Am. A}\ }\textbf {\bibinfo {volume} {20}},\ \bibinfo {pages} {347--356} (\bibinfo {year} {2003})}\BibitemShut {NoStop}%
\bibitem [{\citenamefont {Schubert}\ \emph {et~al.}(1996)\citenamefont {Schubert}, \citenamefont {Rheinl\"{a}nder}, \citenamefont {Woollam}, \citenamefont {Johs},\ and\ \citenamefont {Herzinger}}]{Schubert:96}%
  \BibitemOpen
  \bibfield  {author} {\bibinfo {author} {\bibfnamefont {M.}~\bibnamefont {Schubert}}, \bibinfo {author} {\bibfnamefont {B.}~\bibnamefont {Rheinl\"{a}nder}}, \bibinfo {author} {\bibfnamefont {J.~A.}\ \bibnamefont {Woollam}}, \bibinfo {author} {\bibfnamefont {B.}~\bibnamefont {Johs}},\ and\ \bibinfo {author} {\bibfnamefont {C.~M.}\ \bibnamefont {Herzinger}},\ }\bibfield  {title} {\enquote {\bibinfo {title} {Extension of rotating-analyzer ellipsometry to generalized ellipsometry: determination of the dielectric function tensor from uniaxial {TiO2}},}\ }\href {https://doi.org/10.1364/JOSAA.13.000875} {\bibfield  {journal} {\bibinfo  {journal} {J. Opt. Soc. Am. A}\ }\textbf {\bibinfo {volume} {13}},\ \bibinfo {pages} {875--883} (\bibinfo {year} {1996})}\BibitemShut {NoStop}%
\bibitem [{\citenamefont {Knight}\ \emph {et~al.}(2020)\citenamefont {Knight}, \citenamefont {Schöche}, \citenamefont {Kühne}, \citenamefont {Hofmann}, \citenamefont {Darakchieva},\ and\ \citenamefont {Schubert}}]{KnightRSI2020}%
  \BibitemOpen
  \bibfield  {author} {\bibinfo {author} {\bibfnamefont {S.}~\bibnamefont {Knight}}, \bibinfo {author} {\bibfnamefont {S.}~\bibnamefont {Schöche}}, \bibinfo {author} {\bibfnamefont {P.}~\bibnamefont {Kühne}}, \bibinfo {author} {\bibfnamefont {T.}~\bibnamefont {Hofmann}}, \bibinfo {author} {\bibfnamefont {V.}~\bibnamefont {Darakchieva}},\ and\ \bibinfo {author} {\bibfnamefont {M.}~\bibnamefont {Schubert}},\ }\bibfield  {title} {\enquote {\bibinfo {title} {Tunable cavity-enhanced terahertz frequency-domain optical hall effect},}\ }\href@noop {} {\bibfield  {journal} {\bibinfo  {journal} {Rev. Sci. Instr.}\ }\textbf {\bibinfo {volume} {91}},\ \bibinfo {pages} {083903} (\bibinfo {year} {2020})}\BibitemShut {NoStop}%
\bibitem [{VDI()}]{VDI}%
  \BibitemOpen
  \href@noop {} {\enquote {\bibinfo {title} {{Frequency Multipliers (WR and D Series)}},}\ }\bibinfo {howpublished} {\url{https://vadiodes.com/en/frequency-multipliers}},\ \bibinfo {note} {accessed: 2024-08-23}\BibitemShut {NoStop}%
\bibitem [{\citenamefont {Rindert}\ \emph {et~al.}(2024)\citenamefont {Rindert}, \citenamefont {Richter}, \citenamefont {K\"uhne}, \citenamefont {Ruder}, \citenamefont {Darakchieva},\ and\ \citenamefont {Schubert}}]{PhysRevB.110.054413}%
  \BibitemOpen
  \bibfield  {author} {\bibinfo {author} {\bibfnamefont {V.}~\bibnamefont {Rindert}}, \bibinfo {author} {\bibfnamefont {S.}~\bibnamefont {Richter}}, \bibinfo {author} {\bibfnamefont {P.}~\bibnamefont {K\"uhne}}, \bibinfo {author} {\bibfnamefont {A.}~\bibnamefont {Ruder}}, \bibinfo {author} {\bibfnamefont {V.}~\bibnamefont {Darakchieva}},\ and\ \bibinfo {author} {\bibfnamefont {M.}~\bibnamefont {Schubert}},\ }\bibfield  {title} {\enquote {\bibinfo {title} {Bloch equations in terahertz magnetic-resonance ellipsometry},}\ }\href {https://doi.org/10.1103/PhysRevB.110.054413} {\bibfield  {journal} {\bibinfo  {journal} {Phys. Rev. B}\ }\textbf {\bibinfo {volume} {110}},\ \bibinfo {pages} {054413} (\bibinfo {year} {2024})}\BibitemShut {NoStop}%
\bibitem [{\citenamefont {Richter}\ \emph {et~al.}(2024)\citenamefont {Richter}, \citenamefont {Knight}, \citenamefont {Bulancea-Lindvall}, \citenamefont {Mu}, \citenamefont {K\"uhne}, \citenamefont {Stokey}, \citenamefont {Ruder}, \citenamefont {Rindert}, \citenamefont {Iv\'ady}, \citenamefont {Abrikosov}, \citenamefont {Van~de Walle}, \citenamefont {Schubert},\ and\ \citenamefont {Darakchieva}}]{RichterPRBFebGO2024}%
  \BibitemOpen
  \bibfield  {author} {\bibinfo {author} {\bibfnamefont {S.}~\bibnamefont {Richter}}, \bibinfo {author} {\bibfnamefont {S.}~\bibnamefont {Knight}}, \bibinfo {author} {\bibfnamefont {O.}~\bibnamefont {Bulancea-Lindvall}}, \bibinfo {author} {\bibfnamefont {S.}~\bibnamefont {Mu}}, \bibinfo {author} {\bibfnamefont {P.}~\bibnamefont {K\"uhne}}, \bibinfo {author} {\bibfnamefont {M.}~\bibnamefont {Stokey}}, \bibinfo {author} {\bibfnamefont {A.}~\bibnamefont {Ruder}}, \bibinfo {author} {\bibfnamefont {V.}~\bibnamefont {Rindert}}, \bibinfo {author} {\bibfnamefont {V.}~\bibnamefont {Iv\'ady}}, \bibinfo {author} {\bibfnamefont {I.~A.}\ \bibnamefont {Abrikosov}}, \bibinfo {author} {\bibfnamefont {C.~G.}\ \bibnamefont {Van~de Walle}}, \bibinfo {author} {\bibfnamefont {M.}~\bibnamefont {Schubert}},\ and\ \bibinfo {author} {\bibfnamefont {V.}~\bibnamefont {Darakchieva}},\ }\bibfield  {title} {\enquote {\bibinfo {title} {High-field/high-frequency electron spin resonances of {Fe}-doped
  $\ensuremath{\beta}\text{\ensuremath{-}}{Ga}_{2}{O}_{3}$ by terahertz generalized ellipsometry: Monoclinic symmetry effects},}\ }\href {https://doi.org/10.1103/PhysRevB.109.214106} {\bibfield  {journal} {\bibinfo  {journal} {Phys. Rev. B}\ }\textbf {\bibinfo {volume} {109}},\ \bibinfo {pages} {214106} (\bibinfo {year} {2024})}\BibitemShut {NoStop}%
\bibitem [{\citenamefont {Rindert}\ \emph {et~al.}(2025{\natexlab{a}})\citenamefont {Rindert}, \citenamefont {Darakchieva}, \citenamefont {Sarkar},\ and\ \citenamefont {Schubert}}]{rindert2024prl}%
  \BibitemOpen
  \bibfield  {author} {\bibinfo {author} {\bibfnamefont {V.}~\bibnamefont {Rindert}}, \bibinfo {author} {\bibfnamefont {V.}~\bibnamefont {Darakchieva}}, \bibinfo {author} {\bibfnamefont {T.}~\bibnamefont {Sarkar}},\ and\ \bibinfo {author} {\bibfnamefont {M.}~\bibnamefont {Schubert}},\ }\bibfield  {title} {\enquote {\bibinfo {title} {Magnetic {Lyddane-Sachs-Teller} relation},}\ }\href {https://doi.org/10.1103/PhysRevLett.134.086703} {\bibfield  {journal} {\bibinfo  {journal} {Phys. Rev. Lett.}\ }\textbf {\bibinfo {volume} {134}},\ \bibinfo {pages} {086703} (\bibinfo {year} {2025}{\natexlab{a}})}\BibitemShut {NoStop}%
\bibitem [{\citenamefont {Rindert}\ \emph {et~al.}(2025{\natexlab{b}})\citenamefont {Rindert}, \citenamefont {Galazka}, \citenamefont {Schubert},\ and\ \citenamefont {Darakchieva}}]{RindertAPLCrbGO2024}%
  \BibitemOpen
  \bibfield  {author} {\bibinfo {author} {\bibfnamefont {V.}~\bibnamefont {Rindert}}, \bibinfo {author} {\bibfnamefont {Z.}~\bibnamefont {Galazka}}, \bibinfo {author} {\bibfnamefont {M.}~\bibnamefont {Schubert}},\ and\ \bibinfo {author} {\bibfnamefont {V.}~\bibnamefont {Darakchieva}},\ }\bibfield  {title} {\enquote {\bibinfo {title} {High-frequency/high-field electron paramagnetic resonance generalized spectroscopic ellipsometry characterization of {Cr3+} in $\beta$-ga2o3},}\ }\href {https://doi.org/10.1063/5.0255802} {\bibfield  {journal} {\bibinfo  {journal} {Applied Physics Letters}\ }\textbf {\bibinfo {volume} {126}},\ \bibinfo {pages} {082105} (\bibinfo {year} {2025}{\natexlab{b}})},\ \Eprint {https://arxiv.org/abs/https://pubs.aip.org/aip/apl/article-pdf/doi/10.1063/5.0255802/20411739/082105\_1\_5.0255802.pdf} {https://pubs.aip.org/aip/apl/article-pdf/doi/10.1063/5.0255802/20411739/082105\_1\_5.0255802.pdf} \BibitemShut {NoStop}%
\bibitem [{\citenamefont {Ruder}\ \emph {et~al.}(2020)\citenamefont {Ruder}, \citenamefont {Wright}, \citenamefont {Peev}, \citenamefont {Feder}, \citenamefont {Kilic}, \citenamefont {Hilfiker}, \citenamefont {Schubert}, \citenamefont {Herzinger},\ and\ \citenamefont {Schubert}}]{Ruder:20}%
  \BibitemOpen
  \bibfield  {author} {\bibinfo {author} {\bibfnamefont {A.}~\bibnamefont {Ruder}}, \bibinfo {author} {\bibfnamefont {B.}~\bibnamefont {Wright}}, \bibinfo {author} {\bibfnamefont {D.}~\bibnamefont {Peev}}, \bibinfo {author} {\bibfnamefont {R.}~\bibnamefont {Feder}}, \bibinfo {author} {\bibfnamefont {U.}~\bibnamefont {Kilic}}, \bibinfo {author} {\bibfnamefont {M.}~\bibnamefont {Hilfiker}}, \bibinfo {author} {\bibfnamefont {E.}~\bibnamefont {Schubert}}, \bibinfo {author} {\bibfnamefont {C.~M.}\ \bibnamefont {Herzinger}},\ and\ \bibinfo {author} {\bibfnamefont {M.}~\bibnamefont {Schubert}},\ }\bibfield  {title} {\enquote {\bibinfo {title} {Mueller matrix ellipsometer using dual continuously rotating anisotropic mirrors},}\ }\href {https://doi.org/10.1364/OL.398060} {\bibfield  {journal} {\bibinfo  {journal} {Opt. Lett.}\ }\textbf {\bibinfo {volume} {45}},\ \bibinfo {pages} {3541--3544} (\bibinfo {year} {2020})}\BibitemShut {NoStop}%
\bibitem [{\citenamefont {Ruder}\ \emph {et~al.}(2021)\citenamefont {Ruder}, \citenamefont {Wright}, \citenamefont {Feder}, \citenamefont {Kilic}, \citenamefont {Hilfiker}, \citenamefont {Schubert}, \citenamefont {Herzinger},\ and\ \citenamefont {Schubert}}]{Ruder:21}%
  \BibitemOpen
  \bibfield  {author} {\bibinfo {author} {\bibfnamefont {A.}~\bibnamefont {Ruder}}, \bibinfo {author} {\bibfnamefont {B.}~\bibnamefont {Wright}}, \bibinfo {author} {\bibfnamefont {R.}~\bibnamefont {Feder}}, \bibinfo {author} {\bibfnamefont {U.}~\bibnamefont {Kilic}}, \bibinfo {author} {\bibfnamefont {M.}~\bibnamefont {Hilfiker}}, \bibinfo {author} {\bibfnamefont {E.}~\bibnamefont {Schubert}}, \bibinfo {author} {\bibfnamefont {C.~M.}\ \bibnamefont {Herzinger}},\ and\ \bibinfo {author} {\bibfnamefont {M.}~\bibnamefont {Schubert}},\ }\bibfield  {title} {\enquote {\bibinfo {title} {Mueller matrix imaging microscope using dual continuously rotating anisotropic mirrors},}\ }\href {https://doi.org/10.1364/OE.435972} {\bibfield  {journal} {\bibinfo  {journal} {Opt. Express}\ }\textbf {\bibinfo {volume} {29}},\ \bibinfo {pages} {28704--28724} (\bibinfo {year} {2021})}\BibitemShut {NoStop}%
\bibitem [{\citenamefont {Tompkins}\ and\ \citenamefont {Irene}(2005)}]{TOMPKINS2005xv}%
  \BibitemOpen
  \bibfield  {author} {\bibinfo {author} {\bibfnamefont {H.}~\bibnamefont {Tompkins}}\ and\ \bibinfo {author} {\bibfnamefont {E.}~\bibnamefont {Irene}},\ }\enquote {\bibinfo {title} {Handbook of ellipsometry},}\ \ (\bibinfo  {publisher} {William Andrew Publishing},\ \bibinfo {address} {Norwich, NY},\ \bibinfo {year} {2005})\BibitemShut {NoStop}%
\bibitem [{\citenamefont {Schubert}, \citenamefont {Tiwald},\ and\ \citenamefont {Herzinger}(2000)}]{PhysRevB.61.8187}%
  \BibitemOpen
  \bibfield  {author} {\bibinfo {author} {\bibfnamefont {M.}~\bibnamefont {Schubert}}, \bibinfo {author} {\bibfnamefont {T.~E.}\ \bibnamefont {Tiwald}},\ and\ \bibinfo {author} {\bibfnamefont {C.~M.}\ \bibnamefont {Herzinger}},\ }\bibfield  {title} {\enquote {\bibinfo {title} {Infrared dielectric anisotropy and phonon modes of sapphire},}\ }\href {https://doi.org/10.1103/PhysRevB.61.8187} {\bibfield  {journal} {\bibinfo  {journal} {Phys. Rev. B}\ }\textbf {\bibinfo {volume} {61}},\ \bibinfo {pages} {8187--8201} (\bibinfo {year} {2000})}\BibitemShut {NoStop}%
\bibitem [{\citenamefont {Schubert}(1996)}]{PhysRevB.53.4265}%
  \BibitemOpen
  \bibfield  {author} {\bibinfo {author} {\bibfnamefont {M.}~\bibnamefont {Schubert}},\ }\bibfield  {title} {\enquote {\bibinfo {title} {Polarization-dependent optical parameters of arbitrarily anisotropic homogeneous layered systems},}\ }\href {https://doi.org/10.1103/PhysRevB.53.4265} {\bibfield  {journal} {\bibinfo  {journal} {Phys. Rev. B}\ }\textbf {\bibinfo {volume} {53}},\ \bibinfo {pages} {4265--4274} (\bibinfo {year} {1996})}\BibitemShut {NoStop}%
\end{thebibliography}%

\end{document}